\documentclass[nofootinbib,twocolumn,showpacs,preprintnumbers,amsmath,amssymb,floatfix]{revtex4}

\usepackage{graphicx}
\usepackage{dcolumn}
\usepackage{bm}
\usepackage{color}

\newcommand{\beq}{\begin{equation}}
\newcommand{\eeq}{\end{equation}}
\newcommand{\bey}{\begin{eqnarray}}
\newcommand{\eey}{\end{eqnarray}}

\def\beq{\begin{eqnarray}}
\def\eeq{\end{eqnarray}}

\def\({\left(}
\def\){\right)}

\def\mpl{M_{\rm Pl}}

\def\lsim{\mathrel{\mathstrut\smash{\ooalign{\raise2.5pt\hbox{$<$}\cr\lower2.5pt\hbox{$\sim$}}}}}
\def\gsim{\mathrel{\mathstrut\smash{\ooalign{\raise2.5pt\hbox{$>$}\cr\lower2.5pt\hbox{$\sim$}}}}}

\begin{document}

\preprint{APS/123-QED}

\title{Phenomenological consequences of superfluid dark matter \\ with baryon-phonon coupling}

\author{Lasha Berezhiani} \affiliation{Max-Planck-Institut f\"ur Physik, F\"ohringer Ring 6, 80805 M\"unchen, Germany} 

\author{Benoit Famaey} \affiliation{Universit\'e de Strasbourg, CNRS UMR 7550, Observatoire astronomique de Strasbourg, 11 rue de l'Universit\'e, F-67000 Strasbourg, France}

\author{Justin Khoury} \affiliation{Center for Particle Cosmology, Department of Physics and Astronomy, University of Pennsylvania, Philadelphia PA 19104, USA}%

\date{\today}

\begin{abstract}
Recently, a new form of dark matter has been suggested to naturally reproduce the empirically successful aspects of Milgrom's law in galaxies. The dark matter particle candidates are axion-like, with masses of order~eV and strong self-interactions. They Bose-Einstein condense into a superfluid phase in the central regions of galaxy halos. The superfluid phonon excitations in turn couple to baryons and mediate an additional long-range force. For a suitable choice of the superfluid equation of state, this force can mimic Milgrom's law. In this paper we develop in detail some of the main phenomenological consequences of such a formalism, by revisiting the expected dark matter halo profile in the presence of an extended baryon distribution. In particular, we show how rotation curves of both high and low surface brightness galaxies can be reproduced, with a slightly rising rotation curve at large radii in massive high surface brightness galaxies, thus subtly different from Milgrom's law. We finally point out other expected differences with Milgrom's law, in particular in dwarf spheroidal satellite galaxies, tidal dwarf galaxies, and globular clusters, whose Milgromian or Newtonian behavior depends on the position with respect to the superfluid core of the host galaxy. We also expect ultra-diffuse galaxies within galaxy clusters to have velocities slightly above the baryonic Tully-Fisher relation. Finally, we note that, in this framework, photons and gravitons follow the same geodesics, and that galaxy-galaxy lensing, probing larger distances within galaxy halos than rotation curves, should follow predictions closer to the standard cosmological model than those of Milgrom's law.
\end{abstract}

\maketitle

\section{Introduction}

The standard $\Lambda$ Cold Dark Matter ($\Lambda$CDM) cosmological model provides an excellent fit to the Cosmic Microwave Background (CMB) angular power spectrum, to the large-scale structure matter power spectrum, to the Hubble diagram of Type Ia supernovae, and to the detailed scale of baryonic acoustic oscillations. In the context of simulations of structure formation, it has proven successful down to the scale of galaxy clusters and groups. 

On galactic scales, however, the $\Lambda$CDM model faces some challenges~\cite{Bullock:2017xww}. These include the internal structure of dwarf galaxies in the Local Group~\cite{BoylanKolchin:2011de,toobigtofail2} and possibly beyond~\cite{Papastergis:2014aba}, and the vast planar structures seen around the Milky Way and Andromeda~\cite{Pawlowski:2012vz,Pawlowski:2013kpa,Pawlowski:2013cae,Ibata:2013rh}. Some other issues, notably those related to the frequency, size and pattern speeds of galactic bars, are related to dynamical friction between baryons and a live dark matter (DM) halo, {\it e.g.},~\cite{Debattista:1997bi,Sellwood:2016}. 

Chief among these challenges is arguably the tight correlation between the distribution of baryons and the gravitational acceleration in galaxies~\cite{McGaugh:2016leg,Lelli:2017vgz}. This apparent conspiracy, known as the mass discrepancy acceleration relation (MDAR), states that the acceleration experienced by a baryonic particle (irrespective of whether it is actually due to DM, modified gravity, or both) can be {\it uniquely} predicted from the baryon density profile. At large distances, the MDAR implies the baryonic Tully-Fisher relation (BTFR)~\cite{McGaugh:2011ac}, which relates the total baryonic mass to the asymptotic/flat rotation velocity as $M_{\rm b} \sim V^4_{\rm f}$. This relation holds over five decades in mass, with remarkably small scatter~\cite{LelliBTFR}.  Another scaling relation is the correlation between the central stellar and dynamical surface densities in disk galaxies~\cite{Lelli:2016uea}. 

All of these relations involve a universal acceleration constant $a_0 \sim 10^{-8}{\rm cm}/{\rm s}^{2}$. This constant independently defines $i)$  the zero-point of the BTFR; $ii)$ the transition of the acceleration at which the mass discrepancy between baryonic and total mass appears; $iii)$ the transition surface density between DM-dominated Low Surface Brightness (LSB) galaxies and baryon-dominated High Surface Brightness (HSB) galaxies; $iv)$ the observed maximal baryonic surface density for disk galaxies~\cite{McGaugh:1995ib}; and $v)$ the DM halo central surface density~\cite{Donato:2009ab}.

\subsection{Milgrom's law}

These independent occurrences of $a_0$ in observations can all be summarized by a simple empirical law proposed by Milgrom over thirty years ago~\cite{Milgrom:1983ca,Milgrom:1983pn,Milgrom:1983zz}. Milgrom's law states that the total gravitational attraction $a$ is approximately $a_{\rm N}$ in the regime $a_{\rm N}\gg a_0$, and approaches the geometric mean $\sqrt{a_{\rm N} a_0}$ whenever $a_{\rm N}\ll a_0$, where $a_{\rm N}$ is the usual Newtonian gravitational field generated from the observed distribution of baryonic matter alone. This simple recipe successfully accounts for a wide range of galactic phenomena. It is remarkable that it was suggested by Milgrom well before being corroborated by detailed data: hence, Milgrom made succesful predictions for the behavior of, for instance, LSB galaxies which were not even known to exist at the time. Today, the detailed shape of rotation curves of disk galaxies with vastly different structural parameters is reproduced with uncanny precision by this empirical law. See~\cite{Sanders:2002pf,Famaey:2011kh} for comprehensive reviews.

Although historically proposed as an alternative to DM, hence the name {\it Modified Newtonian Dynamics} (MOND), the {\it empirical} power of Milgrom's formula at fitting galaxy properties is unequivocal, especially in rotationally supported systems. A debate exists, however, as to the possible origin of this law:

\begin{enumerate}

\item The conservative viewpoint is that Milgrom's law is an emergent phenomenon, linked to efficient feedback effects on galaxy scales. This possibility faces a number of challenges. Firstly, semi-empirical models with cored DM halos can reproduce the observed slope and normalization of the BTFR, but not yet the small scatter~\cite{LelliBTFR,DiCintio:2015eeq,Desmond:2017}. Secondly, the diversity of shapes of rotation curves at a given maximal circular velocity remains a puzzle in $\Lambda$CDM~\citep{Oman:2015xda} (though see also~\cite{Read:2016}). Finally, the observation that the central slope of the rotation curve correlates instead with the baryonic surface density means that feedback processes should be more efficient at removing DM from the central regions of galaxies with lower surface density, which are often more gas-dominated. Hence, the feedback efficiency would have to increase with decreasing star formation rate~\cite{McGaugh:2011ac}, and be tightly anti-correlated with baryonic surface density (or correlated with scale-length at a given mass scale). 

More generally, the feedback explanation could present a fine-tuning problem for galaxies that have had a complex formation history, as all effects not taken into account when dealing with internal feedback processes~\citep{Read:2016}, such as the effect of environment and the merger history of each individual galaxy, must then be precisely tuned with the internal feedback in order to give precisely the observed relation~\cite{Famaey:2011kh}.
 
\item At the other end of the spectrum is the viewpoint that Milgrom's law represents a fundamental modification of gravity~({\it e.g.,}~\cite{Bekenstein:1984tv,Bekenstein:2004ne,Milgrom:2009bi}). See~\cite{Famaey:2011kh,Bruneton:2007si} for reviews of such theories. By now this possibility seems however rather unlikely, given the observational evidence for DM behaving as a collisionless fluid on cosmological and cluster scales. Indeed, Milgrom's law does not work on galaxy cluster scales~\cite{Gerbal:1992,Sanders:1999,Clowe:2003tk,Clowe:2006eq,Angus:2006ev,Angus:2008}, and modified gravity theories cannot {\it a priori} explain the angular power spectrum of the CMB without additional DM or unrealistically massive ordinary neutrinos, with a mass above the terrestrial experimental constraints~\cite{Skordis:2005xk}. Such theories are also highly constrained in the Solar System~\cite{Hees:2014kta,Hees:2015bna} (though see~\cite{Babichev:2011kq}). Moreover, Milgrom's law has some problems at subgalactic scales~({\it e.g.,}~\cite{Ibata:2011ri}). 

\item A middle-ground interpretation is that Milgrom's law is telling us something about the {\it fundamental nature} of DM. Indeed, while the existence of a new degree of freedom acting as a pressureless and dissipationless fluid is on firm grounds within the linear regime in cosmology, the exact nature of this dark fluid is far from known. 

Already in the context of self-interacting DM (SIDM)~\cite{Spergel:1999mh}, some recent encouraging results have shown how underdense halos could indeed be associated with extended baryonic disks~\cite{Kamada:2016euw,Creasey:2016jaq}, in line with observations. However, getting a BTFR as tight as observed, as well as a MDAR at all radii remains challenging. A more radical possibility is Modified DM, inspired by gravitational thermodynamics~\cite{Ho:2010ca,Ho:2011xc,Ho:2012ar,Edmonds:2017zhg}.

Our viewpoint here is that the correlation between DM and visible matter embodied in Milgrom's law could be the result of a novel interaction between DM and baryons~\cite{Blanchet:2006yt,Blanchet:2008fj,Khoury:2014tka,Blanchet:2015sra,Blanchet:2015bia,Salucci:2017cet}. Indeed, this could naturally avoid the problems of pure-CDM and pure-MOND, while in principle also reproducing the observed phenomenology on galactic scales.

\end{enumerate}

\begin{table*}
\caption{\label{observationssummary}Summary of observational consequences of superfluid DM}
\begin{ruledtabular}
\begin{tabular}{ll}
System & Behavior\\
\hline
\textbf{Rotating Systems} & \\
    Solar system & Newtonian \\
    Galaxy rotation curve shapes & MOND (+ small DM component making HSB curves rise)\\
    Baryonic Tully--Fisher Relation & MOND for rotation curves (but particle DM for lensing) \\
    Bars and spiral structure in galaxies & MOND  \\
  \textbf{Interacting Galaxies} & \\
    Dynamical friction & Absent in superfluid core \\
    Tidal dwarf galaxies & Newtonian when outside of superfluid core \\
  \textbf{Spheroidal Systems} & \\
    Star clusters & MOND with EFE inside galaxy host core --- Newton outside of core \\
    Dwarf Spheroidals & MOND with EFE inside galaxy host core --- MOND+DM outside of core \\
    Clusters of Galaxies & Mostly particle DM (for both dynamics and lensing) \\
    Ultra-diffuse galaxies & MOND without EFE outside of cluster core \\
  \textbf{Galaxy-galaxy lensing} & Driven by DM enveloppe $\implies$ not MOND \\
    \textbf{Gravitational wave observations} & As in General Relativity\\
\end{tabular}
\end{ruledtabular}
\end{table*}

\subsection{Superfluid dark matter}

Recently, in~\cite{Berezhiani1,Berezhiani2}, we proposed a novel theory of DM superfluidity, inspired by recent developments in cold atom physics. See also~\cite{Khoury:2016ehj,Khoury:2016egg,Hodson:2016rck}. The framework marries the phenomenological success of the $\Lambda$CDM model on cosmological scales with that of Milgrom's empirical law~\citep{Milgrom:1983ca} at fitting galaxy properties. The DM and MOND components have a common origin, representing different phases of a single underlying substance. 
This is achieved through superfluidity.

The DM candidate proposed in~\cite{Berezhiani1,Berezhiani2} consists of axion-like particles with sufficiently strong self-interactions ($\sigma/m \; \lower .75ex \hbox{$\sim$} \llap{\raise .27ex \hbox{$>$}} \; 0.1~{\rm cm}^2/{\rm g}$, where $\sigma$ is the self-interaction cross-section and $m$ the particle mass) such that they thermalize in galaxies. With $m\sim {\rm eV}$, their de Broglie wavelengths overlap in the (cold and dense enough) central region of galaxies, resulting in Bose-Einstein condensation into a superfluid phase. 

The superfluid nature of DM dramatically changes its macroscopic behavior in galaxies. Instead of behaving as individual collisionless particles,
the DM is more aptly described as collective excitations, in particular phonons. The DM phonons couple to ordinary matter, thereby mediating a
long-range force (beyond Newtonian gravity) between baryons. In contrast with theories that propose to fundamentally modify Newtonian gravity,
in this case the new long-range force mediated by phonons is an {\it emergent} property of the DM superfluid medium. 

For a particular choice of the superfluid equation of state, the resulting phonon effective Lagrangian is similar to the Bekenstein-Milgrom MOND theory~\cite{Bekenstein:1984tv}. Remarkably, as reviewed below in Sec.~II, the desired superfluid phonon effective theory is strikingly similar to that of the Unitary Fermi Gas~\citep{UFGreview,Giorgini:2008zz}, which has generated much excitement in the cold atom community in recent years. 

The aim of this paper is to develop novel distinctive predictions of such a model, especially on galaxy scales.
For this purpose we will derive a realistic superfluid halo density profile, improving on the original derivation of~\cite{Berezhiani1,Berezhiani2} by carefully modeling the superfluid at finite temperature. The density profile of~\cite{Berezhiani1,Berezhiani2} used the mean-field Gross-Pitaevskii approach~\cite{G,P}, which assumes that DM is entirely in the condensed phase. In reality, owing to their velocity dispersion DM particles have a small non-zero temperature in galaxies. As is familiar from liquid helium, a superfluid at finite (sub-critical) temperature is best described phenomenologically as a mixture of two fluids~\cite{tisza,london,landau}: $i)$ the superfluid, which by definition has vanishing viscosity; $ii)$ the ``normal" component, comprised of excitations, which is viscous and carries entropy.

The DM density profile should consist of a superfluid core, with approximately homogeneous density, surrounded by an envelope of DM particles in the normal phase
following a collisionless Navarro-Frenk-White (NFW) density profile~\cite{Navarro:1996gj}. In the present contribution, the transition radius will be estimated in two ways. The first estimate, 
denoted by $R_{\rm T}$, is determined by the density dropping to a value where the interaction rate is too low to maintain local thermal equilibrium. The second estimate, denoted by 
$R_{\rm NFW}$, is the radius at which both density and pressure (or equivalently, velocity dispersion) can be matched continuously to an NFW profile. As we shall demonstrate hereafter, for the range of theory parameters that give reasonable fit to rotation curves, $R_{\rm T}$ and $R_{\rm NFW}$ coincide to within $30\%$. 

A key difference with the original papers~\cite{Berezhiani1,Berezhiani2} is that we now choose parameters such that the superfluid core makes up only a fraction of
the total mass/volume of the halo. For the representative massive disk galaxy described in Sec.~\ref{fitrotationcurves}, for instance, the superfluid core makes up $22\%$ of the total DM mass.
The superfluid core must be large enough to encompass the observed rotation curve, since we rely on the phonon-mediated force to reproduce the rotation curve phenomenology.
But it is small enough that most of the DM halo mass is approximately in collisionless form and hence should be triaxial exactly as $\Lambda$CDM halos and consistent with observations~\cite{Clampitt:2015wea}.

A by-product of having a relatively small superfluid core is that most of the gravitational lensing signal will come from the NFW envelope. As a corollary, to reproduce galaxy-galaxy lensing statistics we need {\it not} assume any non-minimal coupling between DM superfluid phonons and photons. Both photons and gravitons travel at the speed of light along the same geodesics, consistent with the tight constraint recently established by the neutron star merger~GW170817~\cite{TheLIGOScientific:2017qsa}. In contrast, some DM-free relativistic completions of MOND, such as~\cite{Bekenstein:2004ne}, rely on non-minimal photon couplings and are now ruled out by GW170817~\cite{Boran:2017rdn}. Potentially measurable effects of non-vanishing phonon sound-speed in the context of future gravitational wave experiments were discussed in~\cite{Cai:2017buj}.

After a few general constraints on the theory parameters in Secs.~\ref{haloprofile} and~\ref{massbound}, the main focus of this paper will be the explicit rotation curve fitting of disk galaxies in Secs.~\ref{algorithm}--\ref{fitrotationcurves}. For this purpose, we will illustrate the method with two representative LSB and HSB disk galaxies (Sec.~\ref{fitrotationcurves}). Given the lack of symmetry, an exact solution for the superfluid/phonon profile would be a daunting task. Instead we will adopt the following hybrid approximation method.

In general the acceleration $\vec{a}$ on a test baryon particle receives three contributions within the superfluid region:
\beq
\vec{a} =\vec{a}_{\rm b} +\vec{a}_{\rm DM}+\vec{a}_{\rm phonon}\,.
\eeq
We calculate the actual Newtonian acceleration generated by the baryons, $\vec{a}_{\rm b}^{\rm actual}$, exactly from the actual non-spherical baryon density. The second term is the gravitational acceleration from the superfluid core, which we derive from a spherically symmetric approximation. The phonon-mediated 
acceleration $\vec{a}_{\rm phonon}$ is sourced by the actual non-spherical baryon density distribution, 
but also depends indirectly on the Newtonian gravitational potential $\Phi$, which we approximate as spherically symmetric for this derivation. As we will show in Sec.~\ref{fitrotationcurves}, the superfluid model offers a satisfactory fit of the rotation curve in both the LSB and HSB cases. 

In Secs.~\ref{clusters}--\ref{DF}, we will describe some of the observational consequences of DM superfluidity for other systems, including galaxy clusters, dwarf satellite galaxies, globular clusters and ultra-diffuse galaxies residing in clusters. A key distinction compared to the MOND predictions for
satellites, in particular, is that the MONDian force in our case is mediated by superfluid phonons and therefore requires superfluidity. In particular, 
the phonon-mediated force between two bodies only applies {\it if both bodies reside within the superfluid region}. If one body is inside
while the other is outside, the only force acting on them is gravity.

The same applies to the so-called external field effect (EFE), an essential aspect of MOND phenomenology. Consider a subsystem with low internal acceleration $a_{\rm int} \ll a_0$ in the presence of a large homogeneous external acceleration $a_{\rm ext}\gg a_0$. In General Relativity (GR), we know from the equivalence principle that $a_{\rm ext}$ has
no physical consequence and can be removed by moving to the freely-falling elevator. In MOND, however, the background acceleration is physical and renders the
subsystem Newtonian.

The EFE is an example of a more general phenomenon in scalar field theories known as {\it kinetic screening}~\cite{Babichev:2009ee,Brax:2012jr,Burrage:2014uwa}.
In theories with gradient interactions, non-linearities in the scalar field gradient --- the scalar acceleration ---  can result in the suppression of the scalar field effects
and the local recovery of standard gravity. See~\cite{Joyce:2014kja} for a review. 

The key difference in the superfluid context is that the EFE is the result of phonon non-linearities, and therefore only applies within the superfluid core. 
In particular, close-by globular clusters, such as Pal~5~\cite{Thomas:2017}, and satellite galaxies residing within the superfluid core of the Milky Way should follow MOND predictions with the EFE. On the other hand, globular clusters at large distances from the Milky Way, such as NGC~2419~\cite{Ibata:2011ri} or Pal~14~\cite{Jordi:2009aw}, are expected to be Newtonian as long as they do not harbor their own DM halo. The same applies to tidal dwarf galaxies resulting from the interaction of massive spiral galaxies. Those are not expected to harbor a significant DM halo, and thus should be Newtonian~\cite{Lelli:2015rba,Lelli:2015yle} as long as they are located outside the superfluid core of their host.

As a preview of the main results of this paper, Table~\ref{observationssummary} summarizes the observational consequences of superfluid DM for various systems.

\section{Effective Description of Superfluid DM}
\label{SFDMreview}

In field theory language, an (abelian) superfluid is described by the theory of  a spontaneously broken global $U(1)$ symmetry, in a state of finite $U(1)$ charge density. At low energy the relevant degree of freedom is the Goldstone boson for the broken symmetry --- the phonon field $\phi$. The $U(1)$ symmetry acts non-linearly on $\phi$ as a shift symmetry, $\phi \rightarrow \phi + c$. Furthermore, in the non-relativistic regime the theory should be Galilean invariant (ignoring gravity). At finite chemical potential $\mu$, the most general effective theory at leading order in derivatives consistent with these symmetries is~\cite{Greiter:1989qb,Son:2002zn} 
\beq
{\cal L}_{T=0} = P(X)\,,
\label{LT=0}
\eeq
where $X=\mu-m\Phi +\dot{\phi}-(\vec{\nabla}\phi)^2/2m$. Here $m$ is the particle mass and $\Phi$ the Newtonian gravitational potential. The conjecture of~\cite{Berezhiani1,Berezhiani2} is that the DM superfluid phonons are governed by the MOND Lagrangian (see~\cite{Famaey:2011kh,Bruneton:2007si})
\begin{equation}
{\cal L}_{{\rm DM}, \,T =  0} = \frac{2\Lambda (2m)^{3/2}}{3} X\sqrt{|X|} \,.
\label{PMOND}
\end{equation}
Remarkably,~\eqref{PMOND} is strikingly similar to that of the Unitary Fermi Gas (UFG)~\cite{UFGreview,Giorgini:2008zz}, which has generated much excitement in the cold atom community in recent years. Indeed, the fractional power of $X$ would be strange if~(\ref{PMOND}) described a fundamental scalar field. As a theory of phonons, however, the power determines the superfluid equation of state, and fractional powers are not uncommon. Indeed, the effective field theory for the UFG superfluid is ${\cal L} \sim X^{n}$, where $n=5/2$ in 3+1 dimensions and $3/2$ in 2+1 dimensions, and is therefore also non-analytic~\cite{Son:2005rv}. 

To mediate a force between baryons, DM phonons must couple to the baryon density as
\beq
{\cal L}_{\rm int} = \alpha\Lambda \frac{\phi}{M_{\rm Pl}} \rho_{\rm b}\,,
\label{coupling}
\eeq
with $\alpha$ being a constant, $\rho_{\rm b}$ the baryonic density, and $M_{\rm Pl}$ the Planck mass. At zero temperature, the effective theory thus has three parameters: the particle mass $m$, a parameter $\Lambda$ related to the self-interaction strength, and the coupling constant $\alpha$ between phonons and baryons. A fourth parameter of the particles themselves is their self-interaction cross-section $\sigma$ setting the conditions for their thermalization, while a fifth parameter $\beta$ will later be introduced to accommodate for finite-temperature effects.\footnote{Note that we will be working mostly hereafter in natural units, where the gravitational potential ($\propto c^2$) is dimensionless, where mass, energy, and acceleration are expressed in eV, and where length and time are in eV$^{-1}$. When dealing with rotation curves or halo masses, we will of course switch back to physical units. In natural units, $m$ and $\Lambda$ can be expressed in eV, while $\alpha$ and $\beta$ are dimensionless.}

From the superfluid perspective, the form of the coupling term~(\ref{coupling}) is unusual: it breaks the $U(1)$ shift symmetry explicitly (albeit softly), making the phonon $\phi$ a pseudo-Goldstone boson. The origin of this particle non-conserving term could be non-perturbative; or more simply, there could be a soft fundamental coupling between DM and baryons which explicitly breaks the relevant symmetries. One way or the other, this makes our quantum liquid a pseudo-superfluid. In particular, the phonon will acquire a mass via radiative corrections, though it is easy to check that the explicit breaking is soft enough that this has no observable effects on galactic scales.

As mentioned already, we do not assume any non-minimal coupling to photons and/or gravitons. Both electromagnetic and gravitational waves travel at the speed of light, consistent with the recent observations of~GW170817~\cite{TheLIGOScientific:2017qsa}. 

Let us briefly review why the action, composed of the terms~\eqref{PMOND} and~\eqref{coupling}, gives rise to the MOND force law. Assuming a static profile, the phonon equation of motion then is
\beq
\label{phononeqsimpler}
\vec{\nabla}\cdot \left( \frac{(\vec{\nabla}\phi)^2- 2m \hat{\mu}}{\sqrt{(\vec{\nabla}\phi)^2- 2m \hat{\mu}}}\vec{\nabla}\phi \right) =\frac{\alpha\rho_{\rm b}}{2\mpl}\,,
\eeq
where $\hat{\mu} \equiv \mu - m\Phi$. In the limit $(\vec{\nabla}\phi)^2\gg 2m \hat{\mu}$, the solution is, ignoring a homogeneous curl term,
\beq
|\vec{\nabla}\phi| \vec{\nabla}\phi \simeq \alpha M_{\rm Pl}\vec{a}_{\rm b}\,,
\label{gradphi}
\eeq
where $\vec{a}_{\rm b}$ is the Newtonian acceleration due to baryons only. The $\phi$-mediated acceleration that derives
from~\eqref{coupling} is 
\beq
\vec{a}_\phi = \alpha \frac{\Lambda}{M_{\rm Pl}}\vec{\nabla}\phi\,. 
\label{aphigen}
\eeq
Thus~\eqref{gradphi} implies
\beq
a_\phi = \sqrt{\frac{\alpha^3\Lambda^2}{\mpl} a_{\rm b}}\,.
\label{deepMOND}
\eeq
As advocated, this matches the deep-MOND form with critical acceleration
\beq
a_0 = \frac{\alpha^3\Lambda^2}{\mpl}\,.
\label{a0us}
\eeq

In the context of MOND, it has long been established that the best-fit value for $a_0$ from galactic rotation curves is
\beq
a_0^{\rm MOND} \simeq 1.2 \times 10^{-8}~{\rm cm}/{\rm s}^2\,.
\label{a0MOND}
\eeq
In our case, the situation is different for two reasons: $i)$ the deep-MOND form~\eqref{deepMOND} is not exact and only applies in the regime $(\vec{\nabla}\phi)^2\gg 2m \hat{\mu}$;
$ii)$ the total acceleration experienced by baryons includes not only  $\vec{a}_{\rm b}$ and $\vec{a}_\phi$, but also $\vec{a}_{\rm DM}$ --- the Newtonian acceleration from the DM halo itself. 
Because the superfluid core is pressure supported, the DM density profile is approximately homogeneous in the central region of galaxies. Its density is moreover rather low. This should be contrasted with the CDM scenario, where DM-only simulations display a cuspy profile~\cite{Navarro:1996gj}.

The story is further complicated by the fact that perturbations around this zero-temperature, static background are unstable (ghost-like). However it was argued in~\cite{Berezhiani1,Berezhiani2} that this instability can naturally be cured by finite-temperature effects. Indeed, our DM particles in galactic halos have non-zero temperature, owing to their velocity dispersion, hence we expect the zero-temperature Lagrangian~\eqref{PMOND} to receive finite-temperature corrections in galaxies.  One way to understand the instability physically is that, if a galactic halo had zero temperature, then the emergent MOND force would destabilize it. 

As it was originally suggested by Landau, a superfluid at finite temperature phenomenologically behaves as a mixture of two fluids. Only a fraction of the total mass forms a degenerate quantum liquid, hydrodynamically described as a potential flow carrying no entropy. The rest of the mass is stored in thermally excited states, behaving as a regular fluid which carries the entropy of the entire system. The general, finite-temperature effective theory, once again at leading order in derivatives, is~\cite{Nicolis:2011cs}
\beq
{\cal L}_{T\neq 0} = F(X,B,Y)\,.
\label{LfiniteT}
\eeq
This generalizes~\eqref{LT=0} through the dependence on two additional scalar quantities $B$ and $Y$ defined by
\begin{eqnarray}
\nonumber
B &\equiv& \sqrt{{\rm det}\;\partial_\mu\psi^I\partial^\mu\psi^J} \,;\\
Y &\equiv& \hat{\mu} +  \dot{\phi} + \vec{v}\cdot \vec{\nabla}\phi \,,
\label{BY}
\end{eqnarray}
where $\psi^I(\vec{x},t)$, $I = 1,2,3$ and $\vec{v}$ are respectively the Lagrangian coordinates and velocity vector of the normal fluid component. Physically, $B$ measures the
density of the normal fluid, while $Y$ is the scalar product of the fluid velocities.

The general finite-temperature effective theory~\eqref{LfiniteT} is only mildly constrained, and further specification of its form requires first-principle knowledge of the superfluid constituents.
Since a fundamental underlying description of our DM superfluid is still lacking,~\cite{Berezhiani1,Berezhiani2} proceeded empirically and considered various finite-temperature actions leading to
MOND-like static profiles with stable perturbations.  One such form, which in this paper we will take as our working model, is
\beq
\mathcal{L}=\frac{2\Lambda (2m)^{3/2}}{3}X\sqrt{| X-\beta Y  |}-\alpha \frac{\Lambda}{\mpl}\phi\rho_{\rm b}\,,
\label{phononaction}
\eeq
where for completeness we have included the coupling term~\eqref{coupling}. This fiducial action may appear {\it ad hoc}, and simpler finite-temperatures corrections were considered in~\cite{Berezhiani1,Berezhiani2},
but it has the practical advantage of making the algebra straightforward. Here $\beta$ is a dimensionless constant that parametrizes finite-temperature effects. In particular the limit $\beta\rightarrow 0$ recovers the original action~\eqref{PMOND}. In what follows we will fix $\beta=2$ as the fiducial value. 

The temperature dependence of the theory is also welcome for another reason. To obtain an acceptable background cosmology and linear perturbation growth, both $\Lambda$ and $\alpha$ must assume different values cosmologically than in galaxies. 
This is not unreasonable, since parameters of the superfluid effective field theory are expected to depend on temperature, and thus on velocity, most naturally through the ratio of the temperature over the critical temperature for phase-transition $T/T_{\rm c}$. In particular they can assume different values on cosmological scales (where $T/T_{\rm c}\sim 10^{-28}$) than in galaxies (where $T/T_{\rm c}\sim 10^{-6}-10^{-2}$). In~\cite{Berezhiani1,Berezhiani2} it was shown that $\alpha$ must be $\sim10^{-4}$ smaller cosmologically, while $\Lambda$ must be $\sim 10^4$ larger, in order to obtain an acceptable cosmology. Hence, the phonon-baryon coupling \eqref{coupling} can be taken as constant.

In the present work, we perform a more detailed analysis of the rotation curves for a more realistic distribution of baryons in two different types of galaxies. As a result we find that, although subdominant in the central parts of the galaxy, the gravitational force due to the superfluid is not always negligible and can be as large as $\sim 30\%$. To compensate for this effect and reproduce the observed rotation curves, we will use a somewhat lower best-fit value, $a_0 \simeq 0.87\times 10^{-8}~{\rm cm/s^2}$.

Another important point to bear in mind concerns the DM mass distribution throughout the halo. It is well known that galaxy clusters pose a problem for MOND. Therefore in order to avoid the tension between our model and observations on cluster scales, it would be most desirable to avoid having a superfluid halo in clusters altogether. It was argued in~\cite{Berezhiani2} that this is possible if the mass of the DM particles is appropriately chosen, and then partly tested in~\cite{Hodson:2016rck}, albeit with slightly more simplified assumptions than in the following. Our more careful analysis will yield somewhat altered bounds on the parameters $m$, $\Lambda$ and $\alpha$ of the theory.

\section{Back-of-the-Envelope Estimates}
\label{haloprofile}

In this Section we revisit the derivation of the DM-only superfluid halo profile originally given in~\cite{Berezhiani2}, and refine various steps in the argument. As in standard $\Lambda$CDM, we expect virialization of halos to proceed through violent relaxation, a manifestly out-of-equilibrium process that should take DM particles out of any pre-existing superfluid state. Therefore, at virialization the halo is expected to have an NFW profile~\cite{Navarro:1996gj}. This initial profile will then be altered in the central regions of the halo, where DM thermalization and condensation occur.

The final equilibrium profile consists of a superfluid core, within which DM particles interact sufficiently frequently to thermalize and form a Bose Einstein condensate, surrounded by an envelope
describing approximately collisionless particles, which we assume for concreteness to be of the NFW form. For the purpose of the simple analytical estimates of this Section, we will approximate the collisionless envelope as a $\rho\sim r^{-3}$ tail of the NFW profile. In the next Section, we will come back and describe a more realistic matching procedure to the full NFW profile, to be applied using numerical analysis.

The conditions for thermal equilibrium and Bose Einstein condensation both depend on the density and velocity profiles, and therefore on the distance $r$ from the center. 
The radii within which thermalization and condensation take place will be denoted respectively by $R_{\rm T}$ and $R_{\rm degen}$. It is easy to argue, however, that
$R_{\rm T}$ must be smaller than $R_{\rm degen}$. Indeed, otherwise there would be a region $R_{\rm degen} < r < R_{\rm T}$ where thermalization is reached without degeneracy, 
which would in turn require an unacceptably large self-interaction cross section. 

On the other hand, since thermalization is necessary to achieve condensation, it would be inconsistent to assume that DM remains in a coherent BEC state
for $R_{\rm T}< r< R_{\rm degen}$. Therefore, we conclude there is only one radius of interest, $R_{\rm T}$. For $r < R_{\rm T}$, our DM is degenerate and
thermalizes through self-interactions. Within this region our density profile consists of a superfluid core, with approximately homogeneous density.  
As mentioned above, for the purpose of making analytical estimates we assume that at $r = R_{\rm T}$ the profile switches to the NFW fall-off $\rho\sim r^{-3}$,
which extends up to the virial radius. 

Before deriving an expression for $R_{\rm T}$, we should be more precise about how large the self-interaction cross section $\sigma$ (assumed velocity-independent) should be to achieve thermal equilibrium. Thermal equilibrium requires the interaction rate of DM particles to be greater than the inverse of the dynamical time, which we take here as
\beq
t_{\rm dyn} \equiv \frac{r}{v(r)}\,,
\eeq
where $v$ is the characteristic velocity dispersion at $r$. It is natural to compare DM interactions to $t_{\rm dyn}$, since the latter sets the characteristic time scale for the motion of perturbers (such as DM subhalos or halo stars) which can disturb the equilibrium state.  

The local self-interaction rate in a region with density $\rho$ and velocity $v$ is given by
\beq
\Gamma = \frac{\sigma}{m} \mathcal{N}v\rho\,,
\label{rate}
\eeq
where $\mathcal{N}\geq 1$ denotes the Bose degeneracy factor,
\beq
\mathcal{N}=\frac{\rho}{m} \left(\frac{2\pi}{mv}\right)^3\,.
\eeq
We will then require $\Gamma > t^{-1}_{\rm dyn}$ for thermal equilibrium.

In the above, both $v$ and $\rho$ depend on the distance $r$ from the center of the halo. This is a key difference from~\cite{Berezhiani2}, where $v$ and $\rho$ were taken as the virial velocity
and density respectively. In what follows we will assume $\mathcal{N}\gg 1$, consistent with the degeneracy discussion above. It is easy to show {\it a posteriori} that this is justified for galaxies
if $m\sim {\rm eV}$.

We will assume for concreteness that thermalization takes place in the outskirts of the halo where the density distribution is given by the tail
of the NFW profile
\beq
\rho(r) = \rho(R_{200}) \left(\frac{R_{200}}{r}\right)^3\,.
\label{rhotailprofile}
\eeq
In this region, up to slowly varying factors (related to the logarithmic growth of the enclosed mass and the varying ratio of velocity dispersion over circular velocity), the velocity profile roughly falls off in a Keplerian fashion as 
\beq
v\sim v_{200}\sqrt{\frac{R_{200}}{r}}\,.
\eeq
In the above, $R_{200}$ and $v_{200}$ are as usual defined as the radius and velocity at which the mean density is 200 times the present critical density:
\begin{eqnarray}
\nonumber
\rho_{200} &=& 200 \frac{3H_0^2}{8\pi G_{\rm N}} \simeq 1.95\times 10^{-27} \; {\rm g}/{\rm cm}^3\,; \\
\nonumber
R_{200} &=& \left(\frac{3M}{4\pi \rho_{200}}\right)^{1/3}  \simeq 203 \left(\frac{M}{10^{12}M_\odot}\right)^{1/3}{\rm kpc}\,;\\
v_{200} &=& \sqrt{\frac{1}{3} \frac{G_{\rm N}M}{R_{200}}}\simeq 85\left(\frac{M}{10^{12}M_\odot}\right)^{1/3}~{\rm km}/{\rm s}\,,
\label{virialrho}
\end{eqnarray}
where we have assumed $H_0 = 70~{\rm km}{\rm s}^{-1}{\rm Mpc}^{-1}$ for concreteness.

Meanwhile, the quantity $\rho(R_{200})$, on the other hand, is the {\it local density} at $R_{200}$, which is of course less than the {\it mean density}
$\rho_{200}$ within that radius. For instance, for a NFW profile with concentration in the range $5 < c < 15$, one finds $4\;\lsim\; \frac{\rho_{200}}{\rho(R_{200})}\;\lsim\; 6$. 
For the purposes of our back-of-the-envelope estimate, we will assume the central value
\beq
\frac{\rho_{200}}{\rho(R_{200})}  = 5\,.
\eeq

Substituting these expressions, the requirement $\Gamma> t^{-1}_{\rm dyn}$ translates to a bound on the cross section:
\beq
\frac{\sigma}{m} \;\gsim\; \left(\frac{m}{{\rm eV}}\right)^{4} \left(\frac{M}{10^{12}M_\odot}\right)^{2/3}\left(\frac{r}{R_{200}}\right)^{7/2} 0.2~\frac{{\rm cm}^2}{{\rm g}}.\nonumber\\
\label{crosssectionH}
\eeq
It is thus easier to reach equilibrium in the inner regions where the density is high. 

Conversely, for a given $\sigma/m$,~\eqref{crosssectionH} translates to a threshold radius $R_{\rm T}$ within which 
thermal equilibrium is achieved. The result is
\beq 
\nonumber
R_{\rm T}  &\lesssim & 310~{\rm kpc} \left( \frac{m}{\rm eV} \right)^{-8/7} \left( \frac{M}{10^{12}M_\odot}\right)^{1/7}\left(\frac{\sigma /m}{{\rm cm}^2/{\rm g}} \right)^{2/7}.
\\
\label{thermalradius}
\eeq
Note that this bound is a rather insensitive function of $M$, and depends more sensitively on $m$.  

Since we hope to explain galaxy rotation curves using DM superfluidity, the superfluid core should extend at least as far as the last observed point on the rotation curve in each galaxy.
In other words, the viability of our scenario requires $R_{\rm T}$ to be larger than the radius within which the circular motion of stars and gas is observed. For instance, consider a 
Milky Way-like galaxy with $M \simeq 10^{12}\;M_\odot$, whose rotation curve (or the equivalent circular velocity curve derived from stellar kinematics, see~\cite{Xue:2008}) is typically measured out to $\simeq 60~{\rm kpc}$. For such a galaxy, we impose that
\beq
R_{\rm T} > 60~{\rm kpc}\,;\qquad \text{for}~~~M = 10^{12}\;M_\odot\,.
\label{RTboundMW}
\eeq
Using~\eqref{thermalradius} this translates to an upper bound on the mass of the DM particle:
\beq
m\lesssim 4.2 \left( \frac{\sigma /m}{{\rm cm}^2/{\rm g}} \right)^{1/4}~{\rm eV}\,.
\label{mupper}
\eeq
Smaller and less massive galaxies result in a somewhat weaker bound.

The bound~\eqref{mupper} on the DM particle mass is the main result of this Section. It shows that for values of $\sigma/m$ satisfying the merging-cluster bound $\sim 1 \, {\rm cm}^2/{\rm g}$ \cite{Randall:2007ph,Massey:2010nd,Harvey:2015hha,Wittman:2017gxn}, $m$ must be somewhat below 4~eV. The dependence on the cross section is rather weak, however, scaling as the $1/4$ power. It should be mentioned that the upper bound~\eqref{mupper} would be somewhat tighter had we assumed a $\rho\propto r^{-2}$ transition density profile outside the superfluid core, instead of $\rho\propto r^{-3}$.

In practice, for the fiducial parameter values that we will use later to fit rotation curves, we will find that the superfluid cores make up a modest fraction of the total DM mass, ranging between 
$\sim 20\%$ and $\sim 50\%$. This has important consequences for observations. Firstly, since most of the mass is in the approximately collisionless NFW envelope, our halos should be triaxial
near the virial radius, exactly as in $\Lambda$CDM halos and consistent with observations~\cite{Clampitt:2015wea}. 

Secondly, most of the gravitational lensing signal will come from the NFW envelope, particularly for HSB galaxies, hence we need not assume any non-minimal coupling between DM superfluid phonons and photons. If we assume that the relation between stellar mass and halo mass is mostly unaltered with respect to $\Lambda$CDM (we will explicitly show that this leads to acceptable rotation curves in Sec.~\ref{fitrotationcurves}), then galaxy-galaxy lensing statistics would be mostly unaletered too. In particular, if Milgrom's law were valid as a fundamental law out to large distances from the baryonic bulk of galaxies, then one would expect a fine relation between the baryonic mass and the weak lensing signal, of the form $r_{\rm E} \propto \sqrt{M_{\rm b}}$, where $r_{\rm E}$ is the Einstein radius \cite{Tian:2009}. Such a relation is not expected to hold in the superfluid DM context.

\section{Lower bound on halo mass}
\label{massbound}

Simulations indicate that virialization takes place when the average density of a perturbation reaches about two hundred times the mean density. Simply put, the over-density is then sufficient for the perturbation to withstand the Hubble expansion. This fact, combined with the observation that pressure supported halos tend to have lower central densities, raises the caution flag regarding the lightest possible haloes one could have in the model in question (see~\cite{Hui:2016ltb} and references therein).

As argued in Sec.~\ref{haloprofile}, the galactic halo consists of a superfluid core surrounded by the NFW envelope. However, the picture simplifies for low mass galaxies, for
which $\rho_{\rm T}\ll \rho_{200}$. Here, $\rho_{\rm T}$ denotes the DM density at the thermalization radius given by~\eqref{thermalradius}. In this case most of the mass resides in the superfluid phase. Ignoring the contribution from phonon gradients,
the equation of state implied by~\eqref{PMOND} is
\beq
P = \frac{\rho^3}{12\Lambda^2m^6}\,.
\label{eos}
\eeq
In~\cite{Berezhiani2} it was shown that this equation of state, together with the condition of
hydrostatic equilibrium, results in a cored density profile with central density 
\beq
\nonumber
\rho_0 \simeq \left(\frac{M}{10^{12}M_\odot}\right)^{2/5} \left( \frac{\Lambda}{{\rm meV}}\frac{m^3}{{\rm eV}^3}\right)^{6/5}\,6.2\times 10^{-25}~{\rm g}/{\rm cm}^3\,.\\
\label{rhocentral}
\eeq
For consistency, the central density $\rho_0$ should be larger than $\rho_{200}$, given in~\eqref{virialrho}.
(Instead of $\rho_0$,~\cite{Hui:2016ltb} uses the average density within $r_{1/2}$, resulting in a tighter constraint.)
This translates to a minimum halo mass $M_{\rm min}$ for the superfluid scenario
\beq
M_{\rm min} \simeq \left(12\; \frac{\Lambda}{{\rm meV}}\frac{m^3}{{\rm eV}^3}\right)^{-3}\,10^9 M_\odot\,.
\eeq
The superfluid pressure suppresses the formation of halos of mass less than $M_{\rm min}$.  

To be consistent with observations, $M_{\rm min}$ should be sufficiently small to accommodate the
least massive galaxies observed. Ultra-faint dwarf satellites around the Milky Way are estimated to have
virial masses between $10^8$ and $10^9~M_\odot$ (before tidal stripping). Demanding that $M_{\rm min} \;\lsim\; 10^9 M_\odot$ to be generous translates to
\beq
\Lambda m^3\gtrsim 0.08~{\rm meV} \times {\rm eV}^3\,.
\label{lowerboundparam}
\eeq
The combination $\Lambda m^3$, which will be a recurring theme in our analysis, traces back to the
equation of state~\eqref{eos} depending on this combination of parameters. In particular, we will see in
the next Section that rotation curves are controlled precisely by this combination and tend to favor lower
values of $\Lambda m^3$. There is therefore an interesting tension between the ability to form light enough halos,
and obtaining a reasonable fit to rotation curves. In practice we will find that values of $\Lambda m^3$ that barely
satisfy~\eqref{lowerboundparam} result in acceptable fits.

\section{Halo profile: Algorithm}
\label{algorithm}

In this Section we show how to derive the superfluid DM density profile in a galaxy, and the predicted rotation curve, for a given (observed) baryonic density profile. As already discussed, the central parts of the halo are expected to be in the superfluid phase while the outskirts of the halo are in the normal phase. Apart from theory parameters, our density profile will depend ultimately on a single parameter --- the total halo mass. This parameter will in turn fix the central value of the potential and DM density. To go further, we thus need a prescription relating the baryonic mass to the halo mass.

Since cosmological simulations of DM superfluid are yet to be devised, we must speculate somewhat on a realistic scenario, which will have to be backed by simulations. To cover the range of possibilities, we will focus on two radically different scenarios. The first scenario assumes that the DM halo mass function is the same as in $\Lambda$CDM (at least at halo masses above the cutoff), and that the same abundance matching prescriptions can be applied ({\it e.g.},~\cite{Behroozi:2012iw}). This scenario implies very different baryon fractions in galaxies of different stellar masses. The second scenario assumes that a given DM halo can, to a first approximation, be thought of as a microcosm of the whole Universe, and can {\it a priori} be expected to host a fraction of baryons representative of the cosmological ratio. On galaxy scales, we could expect a constant fraction (typically of the order of one half) of those baryons to condense towards the center of the halo to form the visible galaxy, thus assuming a negligible effect of feedback. This second scenario thus implies assuming a constant DM over visible baryons ratio $M_{\rm DM}/M_{\rm b} = 10$. Remarkably, we will show in Sec.~\ref{fitrotationcurves} that those two extreme assumptions have a relatively small influence on the rotation curve predictions.

Let us begin with the superfluid region. The Poisson equation is as usual given by
\beq
\nabla^2\Phi=4\pi G_{\rm N} \left(\rho_{\rm SF}+\rho_{\rm b}\right)\,.
\label{poisson}
\eeq
The baryonic mass density $\rho_{\rm b}$ is prescribed by observations. The superfluid DM mass density $\rho_{\rm SF}$ can readily obtained by differentiating the action~\eqref{phononaction} with respect to the Newtonian gravitational potential $\Phi$:
\beq
\rho_{\rm SF} = \frac{2\sqrt{2}m^{5/2}\Lambda\left( 3(\beta-1)\hat{\mu}+(3-\beta) \frac{(\vec{\nabla} \phi)2}{2m}\right)}{3\sqrt{(\beta-1)\hat{\mu}+\frac{(\vec{\nabla} \phi)2}{2m}}}\,.
\label{rhoDMphonon}
\eeq

We will solve Poisson's equation together with the phonon equation
\beq
\label{phononeq}
\vec{\nabla}\cdot \left( \frac{(\vec{\nabla}\phi)^2+2m\left( \frac{2\beta}{3}-1 \right)\hat{\mu}}{\sqrt{(\vec{\nabla}\phi)^2+2m(\beta-1)\hat{\mu}}}\vec{\nabla}\phi \right)=\frac{\alpha\rho_{\rm b}}{2\mpl}\,.
\eeq
The baryon density is of course not spherically symmetric in general. Nevertheless, if we ignore a homogeneous ``pure curl" solution, then this equation can be readily integrated:
\beq
\label{sphericalphonon}
\frac{(\vec{\nabla}\phi)^2+2m\left( \frac{2\beta}{3}-1 \right)\hat{\mu}}{\sqrt{(\vec{\nabla}\phi)^2+2m(\beta-1)\hat{\mu}}}\vec{\nabla}\phi=\alpha \mpl \vec{a}_{\rm b}\,,
\eeq
where $\vec{a}_{\rm b}$ is the Newtonian acceleration due to baryons only. The approximation of neglecting a curl term has been shown in MOND to be accurate to within $\sim 20\%$~\cite{Brada:1994pk}. 

The upshot of this approximation is that~\eqref{sphericalphonon} implies an algebraic equation for the gradient of the phonon field. Specifically, it can be massaged into a cubic equation for $(\vec{\nabla}\phi)^2$. Only one root of this cubic equation is physical --- it is easy to see that the other two roots are complex at large distances where $(\vec{\nabla}\phi)^2\rightarrow 0$. The explicit expression for the physical solution is complicated and not particularly illuminating. Suffice to say the result is we have now determined the phonon gradient as a function of $\Phi$ and $a_{\rm b}$:
\beq
\vec{\nabla}\phi  = F(\Phi,a_{\rm b}) \vec{a}_{\rm b}\,.
\label{gradphiPhiab}
\eeq
Substituting the resulting $(\vec{\nabla}\phi)^2$ into~\eqref{rhoDMphonon} then gives the superfluid mass density in terms of $\Phi$ and $a_{\rm b}$:
\beq
\rho_{\rm SF} = \rho_{\rm SF}(\Phi,a_{\rm b})\,.
\label{rhoDMsolved}
\eeq
Let us stress again that this result holds for an arbitrary (non spherically-symmetric) baryon distribution. Our only approximation was neglecting a pure-curl homogeneous solution in~\eqref{sphericalphonon}. Given a general $\rho_{\rm b}$, the DM-free Poisson equation $\nabla\cdot \vec{a}_{\rm b} = -4\pi G_{\rm N} \rho_{\rm b}$ can be solved for $\vec{a}_{\rm b}$. In turn, the resulting $\vec{a}_{\rm b}$ can be substituted in~\eqref{rhoDMsolved} to obtain $\rho_{\rm SF}$ as a function of $\Phi$, which remains as the last unknown variable. 

The gravitational potential is of course determined by Poisson's equation~\eqref{poisson}. Upon substituting $\rho_{\rm SF}(\Phi,a_{\rm b})$, this becomes a highly non-linear equation, whose general solution absent any symmetries is complicated to find. To simplify the analysis in the present paper, we will replace {\it the real baryon distribution by a spherical density profile $\tilde{\rho}_{\rm b}(r)$ which gives the same purely-baryonic Newtonian acceleration $\vec{a}_{\rm b}$ in the disk.} The simplified Poisson equation becomes
\beq
\frac{1}{r^2} \frac{\rm d}{{\rm d}r} \left(r^2 \frac{{\rm d}\Phi}{{\rm d}r}\right) = 4\pi G_{\rm N}\Big(\tilde{\rho}_{\rm b}(r) + \rho_{\rm SF}(\Phi(r),a_{\rm b})\Big),
\eeq
where $\tilde{a}_{\rm b} = G_{\rm N} \tilde{M}_{\rm b}(r)/r^2$ is the purely-baryonic Newtonian acceleration derived from the spherical distribution $\tilde{\rho}_{\rm b}(r)$. 

We integrate this equation numerically with two boundary conditions: $i)$ imposing regularity at the origin, ${\rm d}\Phi/{\rm d}r|_{r= 0} = 0$; $ii)$ specifying the value of $\Phi$ at the origin, $\Phi(r = 0)$. The latter determines the size and the mass of the superfluid core. In other words, halo cores of different sizes have different central values of the
gravitational potential. 

Upon matching to the normal-phase NFW envelope, discussed below, the mass of the superfluid core in turn fixes the total mass of the halo.
In practice, when fitting rotation curves (Sec.~\ref{fitrotationcurves}) we will proceed through reverse engineering: we first prescribe the total mass $M_{\rm DM}$ of the halo, either as a fixed factor of the baryonic mass ({\it e.g.}, $M_{\rm DM}=10\;M_{\rm b}$) or to its value determined by the abundance matching prescription in $\Lambda$CDM; and then find numerically the value of $\Phi(r = 0)$ which gives the desired $M_{\rm DM}$. 

\subsection{Matching to NFW}

Let us discuss in more detail our matching procedure to the NFW envelope. For this purpose we will improve on the basic analytical estimates of Sec.~\ref{haloprofile},
where we assumed for simplicity a sharp transition at the thermalization radius $R_{\rm T}$ to a $1/r^3$ NFW tail. The first improvement is that we now match
to a full NFW profile
\beq
\rho_{\rm NFW}(r) = \frac{\rho_{\rm c}}{x(1+x)^2}\,;\qquad x \equiv \frac{r}{r_s}\,.
\eeq
The density $\rho_{\rm c}$ and concentration radius $r_s$ (equivalently, the concentration parameter $c = R_{200}/r_s$) are determined by our matching conditions.

The second improvement pertains to the matching radius. In reality, we do not expect a sharp transition at $R_{\rm T}$. Instead, there should be a transition region
around $R_{\rm T}$ within which the DM is nearly in equilibrium but is incapable of maintaining long range coherence. In the simpler context of self-interacting DM, for instance, recent simulations have shown that the density profile starts to deviate from the NFW profile at a radius within which DM particles had a chance to re-scatter at least once over the lifetime of the halo~\cite{SIDM}.

To shed light on the nature and spatial extent of the transition regime, we introduce a second characteristic radius $R_{\rm NFW}$ as the location where the profile can be consistently matched to an NFW profile. The matching conditions at $R_{\rm NFW}$ are:

\begin{enumerate}

\item Continuity of density: $\rho_{\rm SF}  = \rho_{\rm NFW}$.

\item Continuity of pressure: $P_{\rm SF} = P_{\rm NFW}$.

\end{enumerate}

\noindent To evaluate the second condition, we compute the radial superfluid pressure using $P_{\rm SF}=\mathcal{L}+\frac{\partial\mathcal{L}}{\partial X}\frac{(\partial_r\phi)^2}{m}$, reducing to the following expression for the Lagrangian density~\eqref{phononaction} (with $\beta = 2$)
\beq
P_{\rm SF}= \frac{2(2m)^{3/2}\Lambda}{3}~ \frac{\hat{\mu}\left(\hat{\mu}+\frac{(\vec{\nabla}\phi)^2}{2m} \right)+2 \left(\frac{(\vec{\nabla}\phi)^2}{2m} \right)^2}{ \sqrt{ \frac{(\vec{\nabla}\phi)^2}{2m} + \hat{\mu}}}\,.
\eeq
The NFW pressure simply follows from integrating the hydrostatic equilibrium condition:
\beq
P_{\rm NFW}(R_{\rm NFW})  = \int_{R_{\rm NFW}}^\infty {\rm d}r\, \rho_{\rm NFW}(r) \frac{G_{\rm N}M(r)}{r^2}\,,
\eeq
where $M(r)$ is the total (DM + baryonic) mass enclosed.

Together with the constraint of fixed total mass and these matching conditions we need to impose an additional requirement in order to uniquely fix the NFW profile and determine $R_{\rm NFW}$. In particular, we further require that the concentration parameter thus determined is close to its $\Lambda$CDM value.

Meanwhile, as before the thermalization radius $R_{\rm T}$ is defined as the location where the interaction rate $\Gamma$ given by~\eqref{rate} drops to
$t_{\rm dyn}^{-1}  = \sqrt{G_{\rm N}\rho}$. As another refinement on our earlier estimates, we will now evaluate this self-consistently by keeping tracking of $\Gamma$ and $t_{\rm dyn}$ as we 
integrate the equations of motion numerically from the center. (Note that for this purpose $\Gamma$ correctly includes the degeneracy factor ${\cal N}$
appropriate for the superfluid region, but is otherwise a  ``vacuum" scattering rate, {\it i.e.}, it ignores possible dressing effects from the superfluid medium.)

For the range of theory parameters that give a good fit to rotation curves, we found that $R_{\rm NFW}$ and $R_{\rm T}$ agree to within 30\%. A calculation of the exact profile between these radii is beyond the scope of the present work and is left for the future. But these estimates at least instruct us on the likely spatial extent of the transition region.  

\subsection{Fiducial theory parameters}

The DM density profile is determined by two combinations of theory input parameters. The first combination is $m\left(\frac{\sigma /m}{{\rm cm}^2/{\rm g}}\right)^{-1/4}$, which enters in $\Gamma t_{\rm dyn}$ and hence determines the size of the superfluid region. As fiducial values we will set 
\beq
m = 1~{\rm eV}\,;\qquad \frac{\sigma}{m} = 0.01~\frac{{\rm cm}^2}{\rm g} \,.
\label{workingvalue}
\eeq
These values are optimal for ensuring that the superfluid core is large enough in galaxies to encompass the baryonic disk, while being small enough in clusters not to contradict observations. 

The second combination is $\Lambda m^3$. This controls the superfluid pressure and correspondingly the density profile of the core. Furthermore, as argued in Sec.~\ref{massbound}, this quantity also determines the lowest possible halo mass that can form in our scenario. As a fiducial value we will choose
\beq
\Lambda m^3=0.05~{\rm meV} \times {\rm eV}^3\,.
\label{workingvalue2}
\eeq
Although strictly speaking this violates the bound~\eqref{lowerboundparam} on the lowest mass halo, we will see in Sec.~\ref{fitrotationcurves} that values of $\Lambda m^3$ consistent with~\eqref{lowerboundparam} also give reasonable fits to rotation curves.  

Finally, the coupling constant will be set to $\alpha=5.7$ to allow for our best-fit value of $a_0= 0.87\times 10^{-8}~{\rm cm}/{\rm s}^2$, while we have set from the start $\beta=2$ in the action~\eqref{phononaction}.

\section{Halo profile for a toy baryon distribution} 
\label{MONDcomp} 
 
Before comparing to actual data, it is instructive to calculate our predicted profile and rotation curve for a toy baryonic distribution. 
For this purpose we assume a ``spherical exponential" profile
\beq
\rho_{\rm b}^{\rm toy} (r) = \frac{M_{\rm b}}{8\pi L^3} e^{-r/L}\,,
\label{rhobtoy}
\eeq
where the characteristic scale $L$ plays the role of a radial ``scale length". 
 
\begin{figure}[t]
\centering
\includegraphics[width=3.5in]{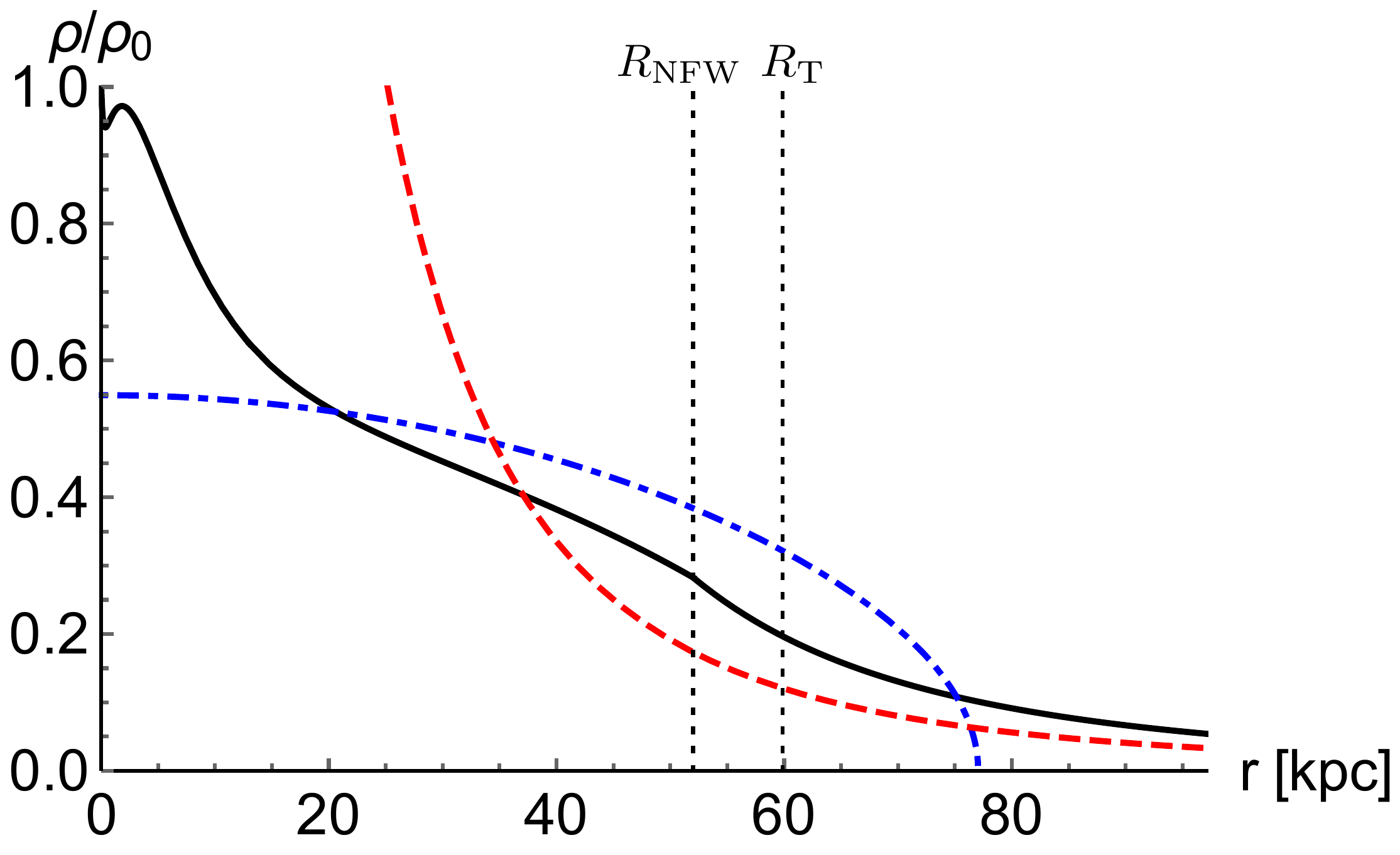}
\caption{\label{DMprofilecompfig} \small Comparison of our DM density profile (solid black curve) for the toy baryon exponential profile~\eqref{rhobtoy}
with $M_{\rm b} = 10^{10}~M_\odot$ and $L = 2~{\rm kpc}$. The parameters are $m = 1~{\rm eV}$, $\Lambda = 0.05~{\rm meV}$ and $M_{\rm DM} = 10\,M_{\rm b} = 10^{11}~M_\odot$.
The transition radii (dashed vertical lines) are $R_{\rm NFW} = 52~{\rm kpc}$ and $R_{\rm T} = 60~{\rm kpc}$. For comparison, also plotted are the ``pure" ($T = 0$, no baryons) superfluid profile (blue dashed-dotted curve), and the NFW profile (red dashed curve) for the same $M_{\rm DM}$ and concentration parameter $c=6.5$.}
\end{figure}
 
Following the algorithm outlined in the previous Section, we can calculate the DM density profile for this exponential baryon distribution once we fix the DM halo mass. Figure~\ref{DMprofilecompfig} illustrates the result (solid black), normalized to the central density, for $M_{\rm b} = 10^{10}\,M_\odot$ and $L = 2$~kpc. 
The value of the central density (or gravitational potential) in this case is fixed to achieve $M_{\rm DM} = 10\;M_{\rm b}$. 
For the theory parameters, we assume the fiducial values~\eqref{workingvalue} and~\eqref{workingvalue2}.
The resulting transition radii (shown as dashed vertical lines on the Figure) are 
\beq
R_{\rm NFW} = 52~{\rm kpc}\,;\qquad R_{\rm T} = 60~{\rm kpc}\,,
\label{toyradii}
\eeq
a difference of $13\%$. Relative to $R_{200} \simeq 97~{\rm kpc}$ for this galaxy,~\eqref{toyradii} implies a significant superfluid core, encompassing $39$\% of the total DM mass. 

The matching conditions then fix the concentration parameter of the NFW envelope $c = 6.5$. For comparison, also shown in the Figure are the ``pure" superfluid profile ($T = 0$, no baryons) derived in~\cite{Berezhiani1,Berezhiani2} (blue dashed-dotted curve), together with the NFW profile (red dashed curve) for the specified mass and concentration. Note that the small
feature within $r\;\lsim\; 2~{\rm kpc}$ is due to the energy density in phonon gradients dominating the superfluid DM mass density.

Using this profile, we can also calculate the radial acceleration on a test baryonic particle. Within the superfluid core this is given by the sum of three contributions:
\beq
a = a_{\rm b}(r)+a_{\rm SF}(r)+a_{\rm phonon}(r)\,.
\eeq
The first term is just the baryonic Newtonian acceleration. The second term is the gravitational acceleration from the superfluid core, $a_{\rm SF} = G_{\rm N}M_{\rm SF}(r)/r^2$.
The third term is the phonon-mediated acceleration~\eqref{aphigen}:
\beq
a_{\rm phonon}(r) =  \alpha \frac{\Lambda}{M_{\rm Pl}} \frac{{\rm d}\phi}{{\rm d}r}\,.
\label{aphononspher}
\eeq
The strength of this term is set by $\alpha$ and $\Lambda$, or equivalently the critical acceleration $a_0$ given by~\eqref{a0us}.

It is instructive to compare our predicted rotation curve with the MOND prediction for a fixed toy baryonic profile. 
For the MOND calculation, we assume the `simple' interpolating function:
\beq
a_{\rm MOND}(r) = \frac{a_{\rm b}(r)}{2} \left(1 + \sqrt{1 + \frac{4a_0^{\rm MOND}}{a_{\rm b}(r)}}\right)\,,
\label{MONDsimple}
\eeq
where $a_{\rm b}(r) = G_{\rm N}M_{\rm b}(r)/r^2$. This is well known to provide good fits to galaxy rotation curves~\cite{Famaey:2005fd,Gentile:2010xt}, as well as elliptical galaxies~\cite{Richtler:2011,Dabringhausen:2016,Chae:2017bhk}.

\begin{figure}[t]
\centering
\includegraphics[width=3.5in]{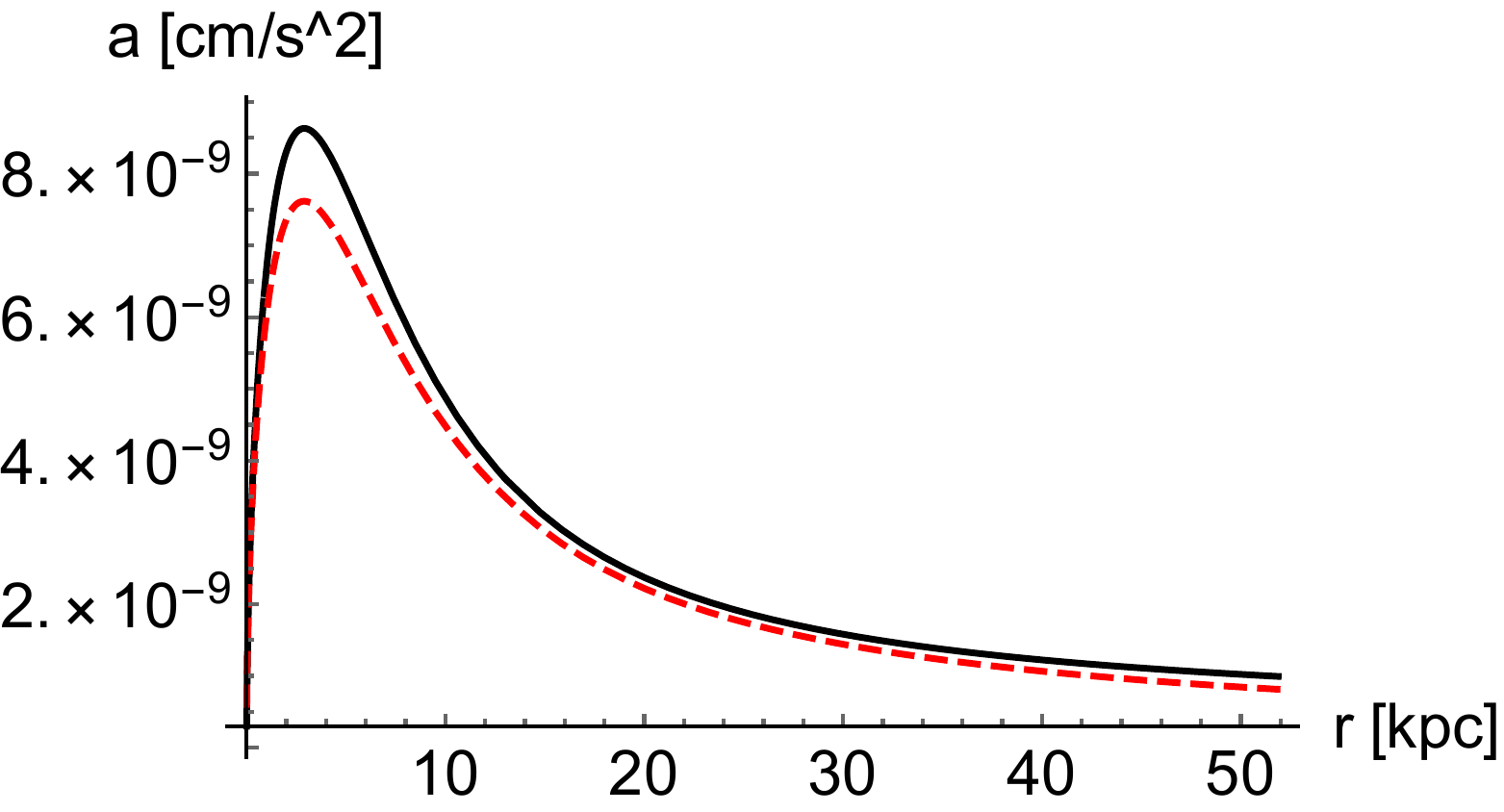}
\caption{\label{MONDcompfig} \small Comparison of our predicted radial acceleration curve (solid black) for the toy baryon exponential profile~\eqref{rhobtoy}. The parameters are the same
as in Fig.~\ref{DMprofilecompfig}. The critical acceleration is $a_0 = 0.87\times 10^{-8}~{\rm cm}/{\rm s}^2$. The MOND prediction (red dashed curve) with the simple interpolating function~\eqref{MONDsimple} and $a_0^{\rm MOND} \simeq 1.2 \times 10^{-8}~{\rm cm}/{\rm s}^2$ for the same baryonic profile is found to agree well (within $25\%$) with our prediction.}
\end{figure}

Figure~\ref{MONDcompfig} compares our predicted radial acceleration curve (solid black curve) with that of the MOND simple interpolating function~\eqref{MONDsimple} (red dashed curve) for the same $M_{\rm b} = 10^{10}\,M_\odot$ and $L = 2$~kpc galaxy as before. All parameters are identical to Fig.~\ref{DMprofilecompfig}. The critical acceleration $a_0$, which normalizes the phonon-mediated force is assumed to be
\beq
a_0 = 0.87\times 10^{-8}~{\rm cm}/{\rm s}^2\,,
\label{bestfit2}
\eeq
which, as we will show in Sec.~\ref{fitrotationcurves}, offers an excellent fit to actual galaxy rotation curves.  
With $\Lambda = 0.05~{\rm meV}$, this corresponds to $\alpha \simeq 5.7$. 

As can be seen from the Figure, we find good agreement with the MOND prediction, to within $\lsim\; 25\%$. 
When fitting galaxy rotation curves in Sec.~\ref{fitrotationcurves}, we will keep our parameters fixed to the
values quoted above.

\section{Galaxy rotation curves}
\label{fitrotationcurves}

In this Section we apply the method outlined above to fit actual galaxy rotation curves.  As it would represent an extremely long exercise to fit a large sample of galaxy rotation curves, we delay such a systematic analysis to future work, and only analyze here, for illustrative purposes, the rotation curves of two disk galaxies, characteristic of the LSB and HSB regime. The LSB galaxy is IC~2574, with total baryonic content $M_{\rm b}\sim 2\times 10^9 M_\odot$. The HSB galaxy is UGC~2953 with $M_{\rm b}\sim 1.6\times 10^{11} M_\odot$. 
 
Our numerical calculation of the superfluid profile and phonon-mediated force in Sec.~\ref{MONDcomp} was performed for a spherically symmetric baryon distribution, whereas the real baryonic distribution is of course far from spherical. To proceed we adopt a hybrid method. As before, the acceleration consists of three terms:
\beq
\vec{a}_{\rm hybrid}=\vec{a}_{\rm b}^{\rm actual} +\vec{a}_{\rm DM}+\vec{a}_{\rm phonon}\,.
\label{hybrid}
\eeq
The different terms are computed as follows:

\begin{enumerate}

\item The baryonic acceleration $\vec{a}_{\rm b}^{\rm actual}$ is the Newtonian acceleration computed for the actual non-spherical baryon density. In other words, $\vec{\nabla}\cdot \vec{a}_{\rm b}^{\rm actual} = - 4\pi G_{\rm N}\rho_{\rm b}^{\rm actual}(\vec{x})$.

\item The DM acceleration $\vec{a}_{\rm DM}$ is the result of our numerical analysis for a simplified spherically-approximated baryon distribution $\rho_{\rm b}^{\rm spherical}$.

\item The phonon-mediated acceleration $\vec{a}_{\rm phonon}$ is then computed using~\eqref{sphericalphonon} sourced by the actual baryon distribution $\vec{a}_{\rm b}^{\rm actual}$, but with the Newtonian potential $\Phi(r)$ taken from our numerical solution to the spherically symmetric problem. Interestingly, despite the non-standard form for the phonon equation~\eqref{phononeqsimpler}, for instance compared to its Bekenstein-Milgrom counterpart~\cite{Bekenstein:1984tv}, we nevertheless found that $a_{\rm phonon}\simeq\sqrt{a_0 a_{\rm b}}$ to within a couple of percent. In other words, the phonon force closely matches the deep-MOND acceleration.
\end{enumerate}

\begin{figure}[t]
\centering
\includegraphics[width=3.5in]{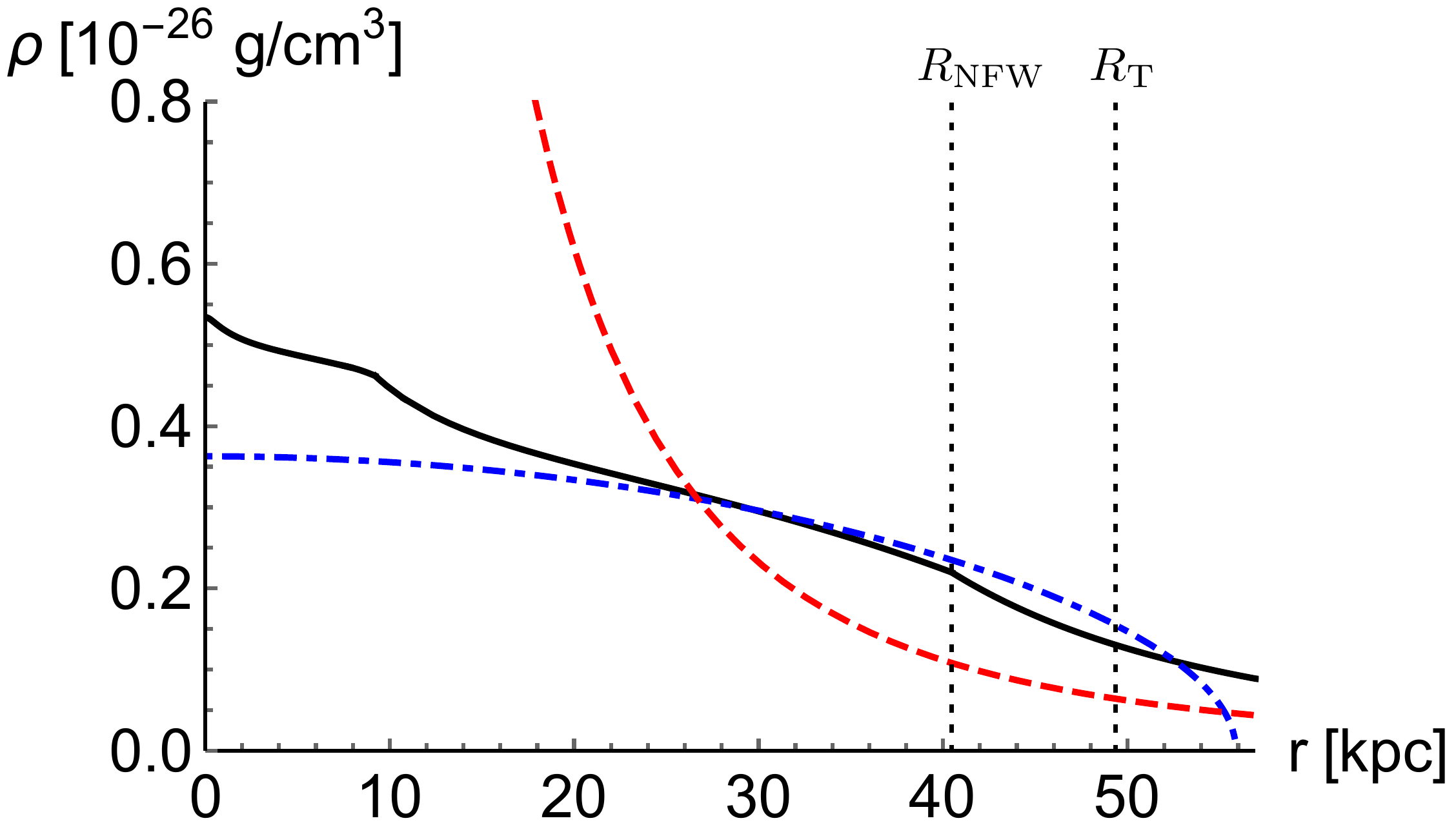}
\caption{\label{DMprofileICfig} \small Superfluid DM density profile (solid black curve) for IC~2574 ($M_{\rm b}\sim 2\times 10^9 M_\odot$). The parameters are $m = 1~{\rm eV}$, $\Lambda = 0.05~{\rm meV}$ and $M_{\rm DM} = 10\,M_{\rm b}$. The transition radii (dashed vertical lines) are $R_{\rm NFW} = 40~{\rm kpc}$ and $R_{\rm T} = 49~{\rm kpc}$, compared to $R_{200} = 57~{\rm kpc}$ for this mass ratio. For the NFW envelope/profile, we have the concentration parameter $c=5.7$. The conventions are the same as in Fig.~\ref{DMprofilecompfig}.}
\end{figure}

\begin{figure}[t]
\centering
\includegraphics[width=3.5in]{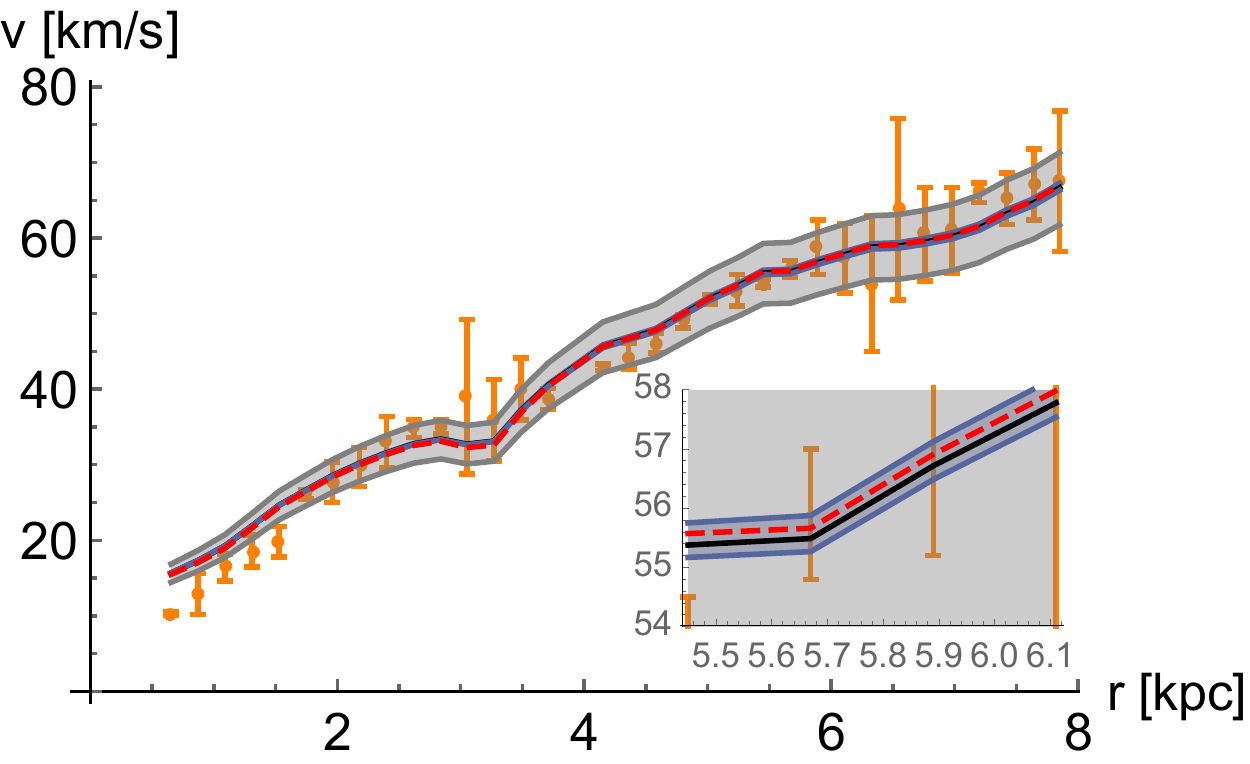}
\caption{\label{ICplot} \small Predicted rotation curve of IC~2574, evaluated using the hybrid method outlined in the text. The \textit{orange points} are the data from~\cite{Lelli:2016zqa} assuming a distance of 3~Mpc~\cite{Tully:2007ue}. The \textit{black curve} is for $M_{\rm DM}=10\;M_{\rm b}$, and assumes the same theory parameter values as Fig.~\ref{DMprofileICfig} and $a_0 = 0.87\times 10^{-8}~{\rm cm}/{\rm s}^2$. The \textit{red dashed curve} assumes the same parameter values except for $M_{\rm DM}\simeq50\;M_{\rm b}$, which is the upper total mass obtained from $\Lambda$CDM abundance matching. The \textit{gray band} corresponds to $a_0\in (0.6,1.2)\times 10^{-8}~{\rm cm}/{\rm s}^2$, while the \textit{light blue band} corresponds to $\Lambda\in (0.02,0.1)~{\rm meV}$, with $M_{\rm DM}=10\; M_{\rm b}$ and other parameters fixed to their fiducial values. The inset zooms in to show the width of the light blue band.}
\end{figure}

\subsection{LSB galaxy (IC~2574)}

Our example of an LSB galaxy is IC~2574 ($M_{\rm b}\sim 2 \times 10^9 M_\odot$). The simplified spherical baryon distribution $\rho_{\rm b}^{\rm spherical}$, necessary to evaluate the DM density profile, was modeled as a constant-density core out to 7.85~kpc, matched to a $1/r^2$ profile out to 9.22~kpc, such that the total mass is $M_{\rm b}\simeq 2\times 10^9 M_\odot$, and that the slope of the baryonic rotation curve is roughly reproduced.

Figure~\ref{DMprofileICfig} shows the DM profile for the fiducial theory parameters~\eqref{workingvalue} and~\eqref{workingvalue2}, 
assuming a constant baryon fraction $M_{\rm DM}=10\,M_{\rm b}$. The resulting transition radii (dashed vertical lines) in this case are 
\beq
R_{\rm NFW} = 40~{\rm kpc}\,;\qquad R_{\rm T} = 49~{\rm kpc}\,,
\label{ICradii}
\eeq
a difference of $18\%$. The matching conditions then fix the concentration radius to $r_s = 10~{\rm kpc}$. 
Relative to $R_{200} \simeq 57~{\rm kpc}$ for this galaxy,~\eqref{ICradii} implies a relatively large superfluid core, encompassing $\simeq 55\%$ of the total DM mass. 

In the abundance matching case~\cite{Behroozi:2012iw}, where $M_{\rm DM} \simeq 51\;M_{\rm b}$ and $R_{200} \simeq 96~{\rm kpc}$, we instead get $R_{\rm NFW}=47~{\rm kpc}$ and $R_{\rm T}=67~{\rm kpc}$; corresponding to $30\%$ difference. In this case the superfluid core containing only $33\%$ of the total DM mass.

Figure~\ref{ICplot} compares the predicted rotation curve, calculated following the hybrid method described above, with the actual data for IC~2574~\citep{Lelli:2016zqa}.

\begin{itemize}

\item The black solid curve corresponds to $M_{\rm DM}=10\;M_{\rm b}$, with theory parameters set to their fiducial values.

\item The red dashed curve (which in this case is barely distinguishable from the black curve) instead assumes $M_{\rm DM}\simeq 51\, M_{\rm b}$, which is consistent with $\Lambda$CDM abundance matching~\cite{Behroozi:2012iw}.

\item The shaded bands illustrate the sensitivity of our result to variations in $\Lambda$ (light blue band, spanning $0.02$ to $0.1~{\rm meV}$)
and $a_0$ (gray band, spanning 0.6 to $1.2 \times 10^{-8}~{\rm cm}/{\rm s}^2$), assuming $M_{\rm DM}=10\;M_{\rm b}$ and keeping other parameters fixed.

\end{itemize}

\noindent The superfluid model recovers the observed rotation curve (orange points) rather well over the specified range of parameters. 

\subsection{HSB galaxy (UGC~2953)}

Our example of a HSB galaxy is UGC~2953 ($M_{\rm b}\simeq 1.6\times 10^{11} M_\odot$). The spherically-approximated baryon distribution $\rho_{\rm b}^{\rm spherical}$ in this case was modeled with two components:

\begin{enumerate}

\item A bulge component modeled by $\rho_{\rm bulge}^{(0)} \left(\frac{L_{\rm bulge}}{r}\right)^\alpha e^{-\left(\frac{r}{L_{\rm bulge}}\right)^\beta}$, 
with $\alpha=1.9$, $\beta=0.8$, $\rho_{\rm bulge}^{(0)}=3.1 \times 10^8 M_\odot /{\rm kpc}^{3}$ and $L_{\rm bulge}=2~{\rm kpc}$.
This profile approximates a deprojected S\'ersic profile~\cite{Mellier}. The corresponding mass of the bulge is $M_{\rm bulge} = 3.5\times 10^{10} M_\odot$. 

\item A sphericized exponential disk $\frac{\Sigma_{\rm disk}}{r}e^{-r/L_{\rm disk}}$, with scale length $L_{\rm disk} = 4~{\rm kpc}$ and central surface density $\Sigma_{\rm disk} = 6.25 \times 10^8 M_\odot/{\rm kpc^2}$. This profile is cut off at $r = 60~{\rm kpc}$, corresponding to the last measured
data point on the rotation curve. The total enclosed mass in the disk is $M_{\rm disk} = 1.25 \times 10^{11} M_\odot$.

\end{enumerate}

\begin{figure}[t]
\centering
\includegraphics[width=3.5in]{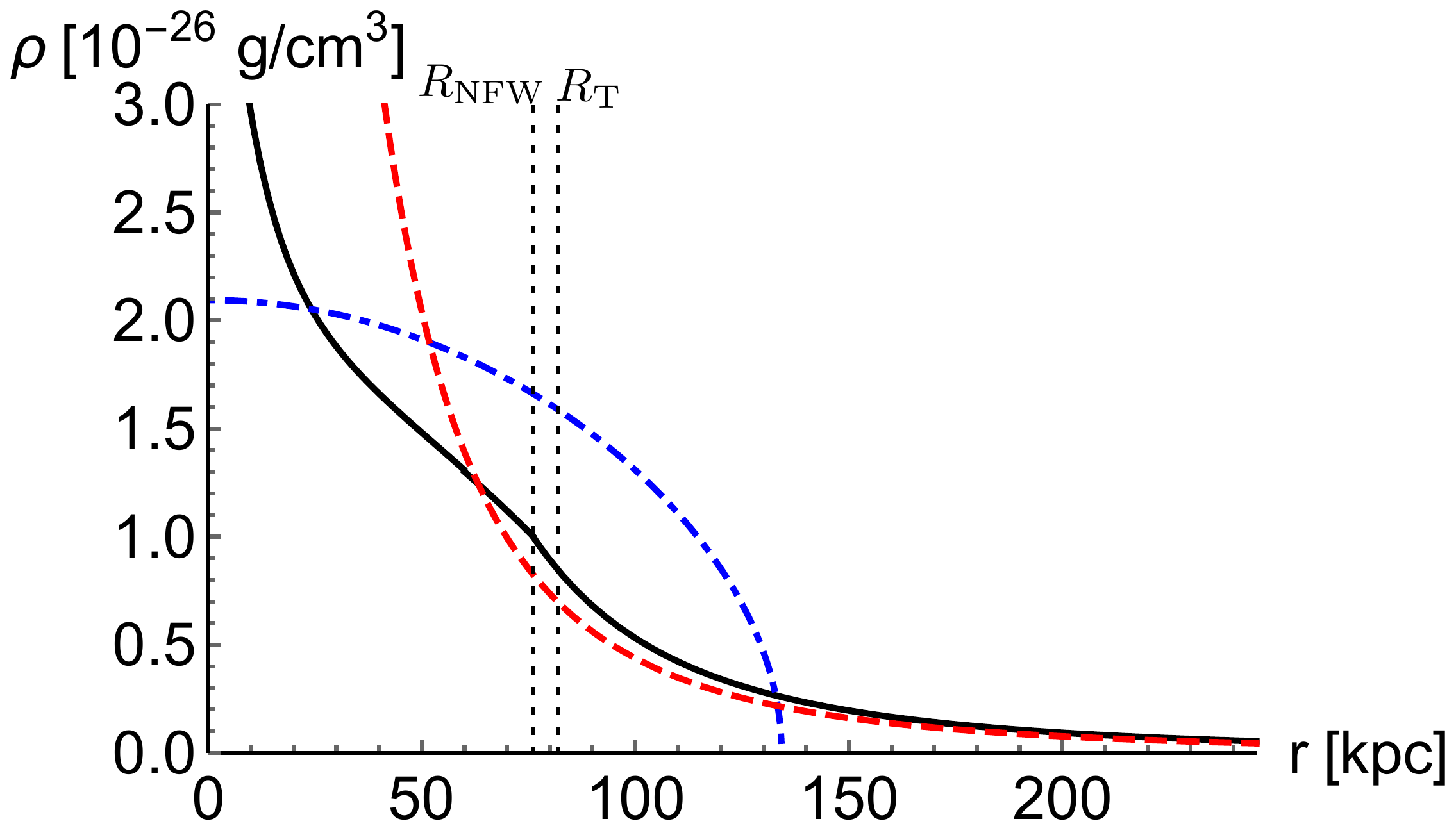}
\caption{\label{DMprofileUGCfig} \small Superfluid DM density profile (solid black curve) for UGC~2953 ($M_{\rm b}\simeq 1.6\times 10^{11} M_\odot$), with $M_{\rm DM} = 10\,M_{\rm b}$. 
Other parameters are the same as in Fig.~\ref{DMprofileICfig}. The transition radii (dashed vertical lines) are $R_{\rm NFW} = 76~{\rm kpc}$ and $R_{\rm T} = 82~{\rm kpc}$, compared to $R_{200} = 245~{\rm kpc}$ for this mass ratio. For the NFW envelope/profile, we have the concentration parameter $c=5.4$. The conventions are the same as in Fig.~\ref{DMprofilecompfig}.}
\end{figure}

\begin{figure}[t]
\centering
\includegraphics[width=3.5in]{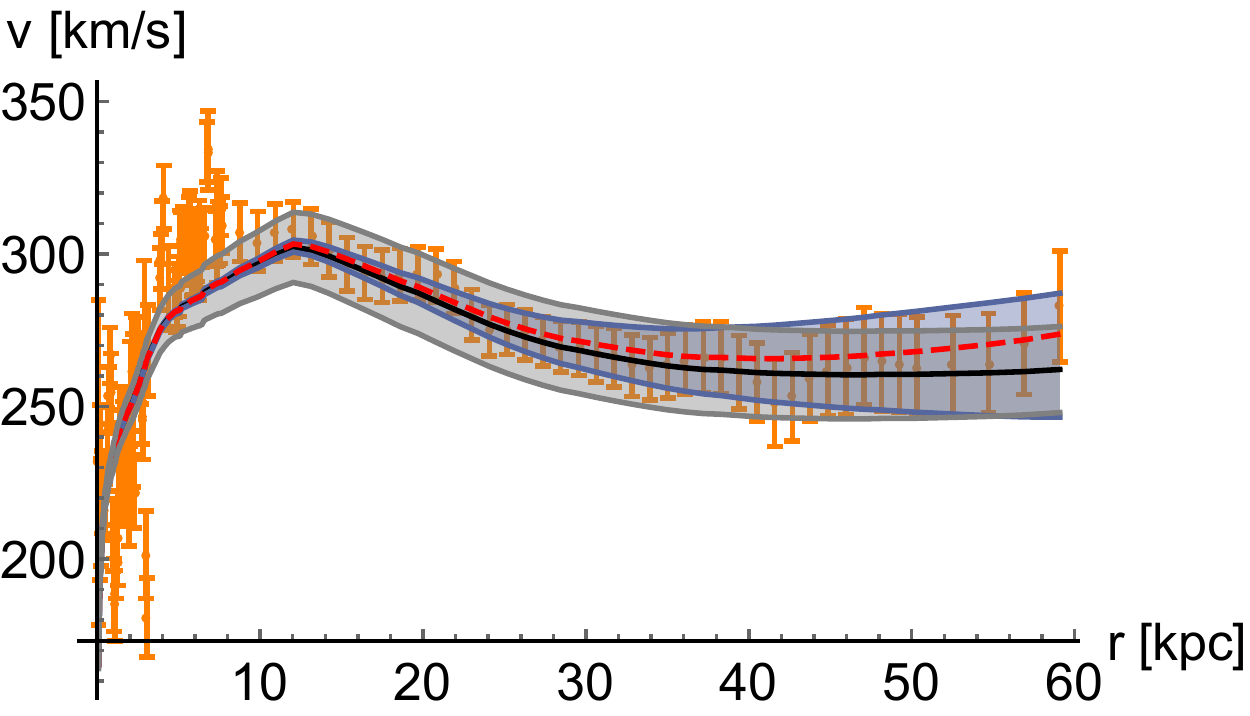}
\vskip 10pt
\caption{\label{UGCplot} \small The rotation curve of UGC~2953, with data ({\it orange points}) taken from~\cite{Noordermeer:2007ux}. 
The parameter values and conventions are the same as in Fig.~\ref{ICplot}. The only difference is the red dashed curve,
which in this case assumes $M_{\rm DM}=65\; M_{\rm b}$, as determined by $\Lambda$CDM abundance matching.}
\end{figure}

Figure~\ref{DMprofileUGCfig} shows the superfluid DM profile for the fiducial theory parameters~\eqref{workingvalue} and~\eqref{workingvalue2}, 
assuming a constant baryon fraction $M_{\rm DM}=10\,M_{\rm b}$. The resulting transition radii (dashed vertical lines) are 
\beq
R_{\rm NFW} = 76~{\rm kpc}\,;\qquad R_{\rm T} = 82~{\rm kpc}\,,
\label{UGCradii}
\eeq
a difference of only $7\%$. The matching conditions then fix the concentration radius to $r_s = 45~{\rm kpc}$. Relative to $R_{200} \simeq 245~{\rm kpc}$ for this galaxy,~\eqref{UGCradii} implies a modest superfluid core, encompassing $22$\% of the total DM mass. 

Figure~\ref{UGCplot} compares the predicted rotation curve with data (orange points) obtained from~\cite{Noordermeer:2007ux}. The parameter values and conventions are the same as for IC~2574. 

The main difference is the red dashed curve, where the total DM mass in this case is set to the $\Lambda$CDM abundance matching value of \cite{Behroozi:2012iw} $M_{\rm DM}=65\;M_{\rm b}$. For this case, we get $R_{\rm NFW}=95~{\rm kpc}$ and $R_{\rm T}=129~{\rm kpc}$, corresponding to a $26\%$ difference. Therefore, the superfluid core, encompassing about $13\%$ of the total DM mass, is significantly smaller than $R_{200}\simeq 446~{\rm kpc}$.
Interestingly, we see a slight increase in the rotation curve at large distances. In contrast with MOND, where the rotation curve becomes flat (or slightly decreases, because of the external field effect~\cite{Hees:2015bna,Haghi:2016jdv}), the slight increase in our case is due to the gravitational force of the superfluid DM itself, on top of the MONDian phonon-mediated force. A precise assessment of the rise of observed rotation curves of HSB galaxies towards the edge of their disk would be a subtle but clean way to distinguish our model from traditional MOND. 

As a final note, we point out that the BTFR in the superfluid DM context is thus tight and MONDian by construction, but that statistical galaxy-galaxy lensing observations, probing larger distances in the halo, should in principle not give the same ``BTFR", but rather give results closer to those expected in $\Lambda$CDM, especially if the stellar to halo mass relation remains the same.

To summarize, the above fiducial galaxies illustrate that the predicted rotation curves of spiral galaxies in the superfluid framework closely resemble
those of MOND, both for LSB and HSB galaxies. The main difference is  limited to the rise of HSB galaxy rotation curves towards the edge of
their observed disks.

\section{Galaxy clusters}
\label{clusters}

In this Section, we make a few basic observations about superfluid DM in galaxy clusters~\cite{Hodson:2016rck}. For this purpose it suffices to revert to the back-of-the-envelope
approach of Sec.~\ref{haloprofile}. 

Because of the larger velocity dispersion in galaxy clusters,  it is easy to show that for the parameter values of interest the Bose enhancement factor $\mathcal{N}$
given in~\eqref{rate} is small at distances comparable to $R_{200}$. Thermalization cannot be achieved at such large distances, thus galaxy clusters are mostly
in the normal (collisionless) phase. Nevertheless, as we go towards the central region and density increases, DM particles can become degenerate and thermalize efficiently. 

Unlike galaxies, clusters are better off, observationally speaking, without a significant superfluid component. 
Taking into account that the thermalized part of the halo is expected to be highly spherical, it is reasonable to demand
\beq
\frac{R_{\rm T}}{R_{200}}\;\lsim\; 0.1\,.
\label{clusterrad}
\eeq
Equivalently, substituting for $R_{200}$ using~\eqref{virialrho} we can write this condition as
\beq
R_{\rm T} \;\lsim\; 200 \left(\frac{M}{10^{15}M_\odot}\right)^{1/3}{\rm kpc}\,.
\label{RTclusterbound}
\eeq

As in Sec.~\ref{haloprofile}, we use the criterion $\Gamma\,t_{\rm dyn} \sim 1$ as an estimator for the size of the superfluid core. 
For such relatively small $R_{\rm T}$, however, we cannot assume that the phase transition happens in the $1/r^3$ tail of the NFW profile. We must instead use an actual NFW profile. On the other hand, because the superfluid region is constrained to be small, we can safely assume that the NFW profile matches the $\Lambda$CDM expectation. Specifically, for a $M=10^{15} M_\odot$ cluster (corresponding to $R_{200} \simeq 2\;{\rm Mpc}$), we assume $c=4$ as concentration parameter, consistent with the mass-concentration relation in $\Lambda$CDM cosmology~\cite{Dutton:2014xda}. 

Using this NFW density profile, and the corresponding velocity profile, we calculate $\Gamma\,t_{\rm dyn}$ as a function of $r$, starting from $R_{200}$ and moving inwards. 
The thermal radius is estimated as in Sec.~\ref{haloprofile} to be the location when $\Gamma\,t_{\rm dyn} = 1$. Substituting the result into~\eqref{clusterrad}, we obtain
a lower bound on the DM mass 
\beq
m\gtrsim 2.7 \left(\frac{\sigma /m}{{\rm cm}^2/{\rm g}} \right)^{1/4}~{\rm eV}\,.
\label{mlower}
\eeq
Combined with the upper bound~\eqref{mupper} derived from galaxies, we obtain the 
allowed range:
\beq
2.7~{\rm eV}\;\lsim\; m\left( \frac{\sigma /m}{{\rm cm}^2/{\rm g}} \right)^{-1/4}\;\lsim\; 4.2~{\rm eV}\,.
\label{mallowedrange}
\eeq
Thus with $\sigma/m$ around the upper limit $\simeq {\rm cm}^2/{\rm g}$ from merging clusters~\cite{Randall:2007ph,Massey:2010nd,Harvey:2015hha,Wittman:2017gxn}, 
$m$ must be in the few eV range.\footnote{An important caveat is that~\eqref{mallowedrange} assumes a constant $\sigma$ for simplicity. Allowing for velocity dependence should
weaken the upper bound, thereby broadening the allowed range.} The fiducial values $m={\rm eV}$ and $\sigma/m= 0.01~{\rm cm}^2/{\rm g}$ adopted in fitting the galaxy rotation curves fall within this range. For instance, they imply $R_{\rm T}\simeq 163~{\rm kpc}$ for the aforementioned cluster, which satisfies~\eqref{RTclusterbound}.

\section{Superfluid External Field Effect}

An important aspect of the MOND phenomenology is the so-called {\it external field effect} (EFE), which is a consequence of the breaking
of the strong equivalence principle. In GR, the equivalence principle tells us that a homogeneous acceleration has
no physical consequence --- it can be removed by moving to the freely-falling elevator. This is not so in MOND. The internal dynamics of
a subsystem are affected by the presence of an external acceleration. See~\cite{Sanders:2002pf,Famaey:2011kh} for reviews.

For instance, consider a subsystem with low internal acceleration ($a_{\rm int} \ll a_0$) in the background of a large external acceleration
($a_{\rm ext}\gg a_0$). Then the EFE implies a screening effect, resulting in approximately Newtonian dynamics in the
subsystem. Consider then the situation of an external-field dominated subsystem ($a_{\rm int} \ll a_{\rm ext}$) immersed in a MONDian background acceleration ($a_{\rm ext}\ll a_0$). In this case, the EFE implies that the subsystem will remain Newtonian, but with an enhanced Newton's constant $G_{\rm eff} \sim \frac{a_0}{a} G_{\rm N}$. 

The EFE is not special to MOND but is an example of a more general phenomenon in scalar field theories with kinetic interactions
known as {\it kinetic screening}~\cite{Babichev:2009ee,Babichev:2011kq,Brax:2012jr,Burrage:2014uwa}. In $P(X)$ theories, of which~\eqref{PMOND} is an example, non-linearities in the 
scalar field triggered by $\vec{\nabla}\phi$ (the acceleration) becoming large can result in the suppression of the scalar field effects
and the local recovery of GR. See~\cite{Joyce:2014kja} for a review.

The main difference in the superfluid context is that the MONDian effective theory~\eqref{PMOND}, or more precisely its finite-temperature
version~\eqref{phononaction}, is only valid within the superfluid core. This implies, first of all, that the phonon-mediated force between two bodies only
applies {\it if both bodies reside within the superfluid region}. If one body is inside while the other is outside, the only force acting on
them is gravity. The same holds true for the EFE --- it is only active {\it within} the superfluid core of the host.

Below we apply these observations to various subsystems in galaxies and in galaxy clusters.

\subsection{Satellite galaxies and globular clusters}

Based on the above discussion, we expect that within the superfluid core of the Milky Way, both globular clusters and satellite galaxies are expected
to follow MOND predictions. In particular, the tidal streams of globular clusters, such as Pal~5, orbiting within the superfluid core should be affected by the EFE just as in traditional MOND~\citep{Thomas:2017}.
The case of dwarf spheroidal galaxies depends on the actual exact size of the superfluid core around the Milky Way. With a core of $\sim 60 \, {\rm kpc}$, most dwarf spheroidals would sit {\it outside} of the core, and should thus {\it not} be affected by the phonons of the superfluid core. In this case, if the dwarf spheroidals are primordial (hence not tidal dwarfs), the expected DM masses from abundance matching being very large, they would mostly be expected to display higher velocities than predicted in MOND, even without the EFE, as observed~\cite{Spergel,Milgrom:1995hz,Angus:2008vs,Lughausen:2014}. Also, globular clusters at large distances from the Milky Way, such as NGC~2419~\cite{Ibata:2011ri} or Pal~14~\cite{Jordi:2009aw}, would be expected to be Newtonian as long as they do not harbor their own DM halo.

In the case of a larger superfluid core of size $\sim 120 \, {\rm kpc}$, dwarfs such as Sextans and Draco would sit well within the core, and would thus need to be explained by a combination of outliers, binaries, extreme anisotropy, and out-of-equilibrium dynamics~\cite{Serra:2009tj,McGWolf:2010}. Globular clusters such as Pal 14 and NGC 2419 would moreover become MONDian. Such a large core would have the advantage, however, to naturally explain the observed velocity dispersion of Crater~II \cite{McGaugh:crater,Caldwell:2017}, which is surprisingly low in the standard context.

The same reasoning applies to tidal dwarf galaxies resulting from the interaction of massive spiral galaxies. Those are not expected to harbor a significant DM halo, and thus should be Newtonian~\cite{Lelli:2015rba,Lelli:2015yle} as long as they are located outside the superfluid core of their host.

Finally, tidal stellar streams of satellite galaxies such as the Sgr dwarf would be expected to display very peculiar features linked to the crossing of the transition radius between the superfluid and normal phase. Once reaching this transition radius, they could be ejected to much larger distances and then come back within the superfluid core on a different orbit. This could leave distinct signatures in tidal streams, such as bifurcations, which would be unexplainable in any other framework. We leave a detailed study of tidal streams to future work.

\subsection{Ultra-diffuse galaxies in galaxy clusters}

As shown in Sec.~\ref{clusters}, galaxy clusters have a relatively small superfluid core, of order $100~{\rm kpc}$ for $m =1~{\rm eV}$. 
Nevertheless, galaxies {\it within} galaxy clusters are still expected to be surrounded by their own superfluid cores. This is interesting, for instance, in view of the recent observations of ultra-diffuse galaxies within clusters~\cite{vanDokkum:2014cea,Koda:2015gwa}, which are observed to be extremely DM dominated~\cite{vanDokkum:2015}. This class of objects poses a problem for MOND because of the EFE~\cite{Milgrom:2015rma,Hodson:2016rck}. Indeed, although ultra-diffuse galaxies have small internal accelerations, and thus
in isolation should be deeply MONDian, the large external acceleration from the galaxy cluster should cause them to be Newtonian, in conflict with observations.
 
In the superfluid case, however, there is no EFE in those galaxies since they are clearly orbiting outside of the superfluid core of the cluster itself. Moreover, these galaxies are very extended, meaning that their baryon fraction must be smaller than the one assumed here for spiral galaxies, in order for their superfluid core to encompass their large observed size. Hence, these LSB galaxies would share some characteristics of the HSB galaxy analyzed above in terms of the mass of their DM halo. This means that they should slightly deviate from the BTFR, with velocities slightly above the BTFR expectation.

\section{Dynamical Friction}
\label{DF}

A distinctive prediction of the superfluid framework is the absence of dynamical friction for subsonic motion within the superfluid region~\cite{Ostriker:1998fa,Hui:2016ltb}.
This may alleviate a number of minor problems for $\Lambda$CDM. For instance, instead of being slowed down by dynamical friction, galactic bars in spiral galaxies
should achieve a nearly constant velocity~\cite{Chandra}, as favored by observations~\cite{Debattista:1997bi}. 

It may also offer a natural explanation to the long-standing puzzle of why the five globular clusters orbiting Fornax have not merged to the center to form a stellar nucleus.
Indeed, in the context of CDM, dynamical friction should have caused the globular clusters to rapidly fall towards the center of Fornax~\cite{Fornaxold,Tremaine}. In reality
Fornax shows no sign of such mergers. See~\cite{Cole,SanchezSalcedo:2006fa} for possible explanations within the classical DM particles context. In our case, the globular clusters should be happily swimming within the superfluid core without dissipation.

As another speculation, the suppression of dynamical friction might shed light on the luminous red galaxy (LRG) two-point correlation function~\cite{Masjedi:2005sc}, which
remains surprisingly featureless ($\xi(r) \sim 1/r^2$) down to small distances where one would expect dynamical friction and mergers to show up. This
observation in fact implies an upper bound for the LRG-LRG merger rate of $\lsim\; 0.6 \times 10^4~{\rm Gyr}^{-1}/{\rm Gpc}^3$~\cite{Masjedi:2005sc}. The suppression of dynamical friction
within the superfluid cores of the LRG may partially explain these observations. 

Two other speculations related to reduced dynamical friction are that it could have interesting consequences on the long term evolution of Hickson compact groups~\citep{Kroupa:2015}, provided that their cores are large enough, as well as on the possible history of the Local Group, allowing for a past Milky Way-Andromeda encounter~\citep{Zhao:2013,Banik:2017}. The latter may require a large superfluid core for the Milky Way. By reducing~\eqref{workingvalue} down to $0.85~{\rm eV}$, which saturates the bound acquired for clusters, one can increase the superfluid core radius to $\sim 100~{\rm kpc}$. One could push it further to even greater values by violating~\eqref{mlower}; however the scattering cross section does not have to be a constant, in fact the mild velocity dependence would suffice to push the core radius well-above $100~{\rm kpc}$ for a HSB galaxy, while simultaneously satisfying~\eqref{mlower} in clusters. Moreover, as we already mentioned, if we set the total DM mass of HSB to the $\Lambda$CDM abundance matching value then the superfluid core radius is already $\sim 100~{\rm kpc}$ without even altering~\eqref{workingvalue}.

In ongoing work, we are currently analyzing quantitatively the effect of dynamical friction in the superfluid context~\cite{BenLasha}. 

\section{Conclusion}

The theory of DM superfluidity proposed in~\cite{Berezhiani1,Berezhiani2} has some commonalities with SIDM and fuzzy DM~\cite{Hu:2000ke,Hui:2016ltb}. All three proposals achieve a cored DM profile in central regions of galaxies, either through interactions or quantum effects, to alleviate existing observational puzzles/tensions with $\Lambda$CDM on
galactic scales. In all three proposals, the DM outside the core is in approximately collisionless form and assumes an NFW profile. Like the quantum pressure of fuzzy DM, the classical
pressure of superfluid DM results in lower central densities and implies a minimal halo mass necessary for collapse and virialization.

The main difference is that DM superfluidity achieves a much larger core, encompassing the entire range of scales probed by rotation curve observations.
Moreover, the superfluid collective excitations (phonons) within the core mediate a long-range force within the core, thereby affecting the dynamics of orbiting baryons and reproducing the MOND phenomenology. Unlike most attempts to modify gravity, there is no
fundamental additional long-range force in the model. Instead the phonon-mediated force is an {\it emergent} phenomenon which requires
the coherence of the underlying superfluid substrate. 

This has important implications for the phenomenological viability of the scenario {\it vis-\`a-vis} solar system tests of gravity. Although typical
accelerations in the solar system are large compared to $a_0$, post-Newtonian tests are sensitive to small corrections to Newtonian gravity
and require further suppression of the new force at short scales. The superfluid framework offers an elegant explanation. As argued in~\cite{Berezhiani1,Berezhiani2},
the large phonon gradient induced by an individual star results in a breakdown of superfluidity in its vicinity. Coherence is lost within the solar system. Individual DM
particles still interact with baryons, but no long-range force can be mediated.

In this paper, using a combination of analytical and numerical arguments, we developed a number of novel distinctive predictions of the superfluid framework. 
For this purpose we derived an approximate finite-temperature density profile, consisting of a superfluid core, with approximately homogeneous density, surrounded by an envelope of normal-phase DM particles following an NFW profile. In this context, we constrained the parameters of the theory from various observational constraints. In particular, we constrained the mass $m$ of the particles such that the superfluid core can be large enough in galaxies, but small enough in clusters. We constrained the combination $\Lambda m^3$ such that massive enough halos could form. 

An important difference of the present analysis compared to the original papers~\cite{Berezhiani1,Berezhiani2} is that the superfluid core makes up only a modest fraction of the entire halo. It is large enough to encompass the observed rotation curves, since the phonon force is critical for reproducing MOND. But it is small enough that most of the mass lies in the approximately collisionless envelope, resulting in triaxial halos near the virial radius.

Since, as we showed, keeping unaltered the relation between stellar mass and halo mass with respect to $\Lambda$CDM abundance matching leads to acceptable rotation curves, galaxy-galaxy lensing statistics should be mostly unaltered too. In particular, contrary to MOND, one does not expect a fine relation $r_{\rm E} \propto \sqrt{M_{\rm b}}$ between the baryonic mass and the weak lensing signal. Also, both photons and gravitons travel at the speed of light along the same geodesics, consistent with the tight constraint recently established by the neutron star merger~GW170817.

The main result of this paper is the explicit rotation curve fitting of two disk galaxies: a representative LSB galaxy IC~2574, and a representative HSB galaxy UGC~2953.
The superfluid model offers an excellent fit in both cases. This is particularly remarkable for IC~2574, which is notoriously difficult to recover in $\Lambda$CDM cosmology
with standard feedback recipes~\cite{Oman:2015xda}. The predicted rotation curve for IC~2574 is phonon-dominated and therefore closely matches the MOND prediction.

The rotation curve for UGC~2953 shows a small but potentially interesting distinguishing feature from traditional MOND. 
In contrast with MOND, where the rotation curve becomes flat (or slightly decreases, because of the EFE~\cite{Hees:2015bna,Haghi:2016jdv}), 
we found a slight rise in the asymptotic velocity. This is due the gravitational pull of the superfluid DM mass itself, which adds to the
phonon acceleration. A precise assessment of the rise of observed rotation curves of HSB galaxies towards the edge of their disk would be a subtle but clean way to distinguish our model from traditional MOND. 

We also explored the observational consequences, albeit at a more qualitative level, for other systems, such as galaxy clusters, dwarf satellites, globular clusters and ultra-diffuse galaxies in clusters. A key difference compared to traditional MOND is that the MONDian force in our case is limited to the superfluid region. Both bodies must be in the superfluid core to experience a long-range phonon force.

Similarly, the EFE (or kinetic screening effect) of MONDian phenomenology only applies within the superfluid core. As a result, globular clusters and satellite galaxies within the superfluid core of the Milky Way should follow MOND predictions. But globular clusters at large distances from the Milky Way should be Newtonian, assuming they do not harbor their own DM halo, while dwarf spheroidals outside of the core might display larger velocities than expected from the BTFR if the usual abundance matching recipes apply. On the other hand, tidal dwarf galaxies resulting from the interaction of massive spiral galaxies should be expected to be devoid of DM and purely Newtonian if they sit outside of the core of their merged host. Finally, ultra-diffuse galaxies in galaxy clusters are immune to the EFE in our case, since they are clearly orbiting outside of the (small) superfluid core of the cluster itself. All these predictions are summarized in Table~I. They will of course have to be backed by numerical simulations of superfluid DM, which are currently underway~\cite{ElderMotaWinther}.

\section*{Acknowledgements}
We thank Lam Hui, Andrey Kravtsov, Tom Lubensky, Jerry Ostriker, Riccardo Penco and HongSheng Zhao for helpful discussions. The work of L.B. was partially supported by US Department of Energy grant DE-SC0007968 at Princeton University. B.F. acknowledges the financial support from the ``Programme Investissements d'Avenir'' (PIA) of the IdEx from the Universit\'e de Strasbourg. J.K. is supported in part by NSF CAREER Award PHY-1145525, US Department of Energy (HEP) Award DE-SC0017804, NASA ATP grant NNX11AI95G, and the Charles E. Kaufman Foundation of the Pittsburgh Foundation.

\end{document}